\documentclass[aps,amsmath,amssymb,amsfonts,twocolumn]{revtex4-2}
\usepackage{graphicx}
\usepackage[colorlinks=true,linkcolor=blue,citecolor=blue, urlcolor=blue]{hyperref}
\usepackage{blindtext}
\usepackage{comment}
\usepackage{multirow}
\usepackage{tabularx}
\usepackage[dvipsnames]{xcolor}
\usepackage{lipsum} 
\usepackage{mathtools}
\usepackage{lineno}
\usepackage{adjustbox}

\usepackage{setspace}
\AtBeginDocument
{	\addtolength\abovedisplayskip{0.20\baselineskip}%
	\addtolength\belowdisplayskip{0.20\baselineskip}%
}

\newcommand{\RNum}[1]{\uppercase\expandafter{\romannumeral #1\relax}}
\newcommand{\RNumLow}[1]{\lowercase\expandafter{\romannumeral #1\relax}}
\newcommand{\q}[1]{``#1''}
\usepackage[normalem]{ulem}
\begin{document}

\title{Reaction dynamics in non-Markovian systems with non-monotonically decaying memory} 

\author{~Artur Bakaev}
\author{~Otto Geburtig}
\author{~Benjamin A. Dalton}
\author{~Roland R. Netz}
\email{corresponding author. rnetz@physik.fu-berlin.de}
\affiliation{Freie Universit\"at Berlin, Fachbereich Physik, 14195 Berlin, Germany}
\date{15. September 2026}

\begin{abstract}
Whereas the effect of monotonically decaying memory kernels on reaction dynamics has been studied extensively, the effect of oscillatory memory, observed for widely different systems, is much less explored. We analyze non-Markovian barrier-crossing dynamics in the presence of a single exponentially decaying oscillatory friction memory kernel by simulations and analytical theory. We identify three scaling regimes: two in which the mean first-passage time $\tau_{\mathrm{MFP}}$ scales quartically and linearly with the memory kernel decay time $\tau_{\mathrm{o}}$, respectively, and a third in which $\tau_{\mathrm{MFP}}$ scales quartically with the oscillation period $\tau_\varphi$ of the memory kernel.
Our results are  relevant to a broad class of many-body dynamical systems such as dihedral isomerization processes and chemical reactions in solvents.
\end{abstract}

\maketitle

\section{Introduction}
\label{sec.1}

Physical, chemical, and biological many-body systems are commonly described in terms of low-dimensional reaction coordinates \cite{kramers1940brownian,visscher1976escape,Chandler78,skinner1978relaxation, van1998remarks}. A prominent example is protein folding, which can be monitored using suitably defined scalar reaction coordinates.\cite{ayaz2021multiMem,dalton2022protein,plotkin1998non,satija2019generalized}. Solving the equations of motion of nonlinear many-body systems analytically is generally impossible, while direct simulations are computationally expensive \cite{ayaz2021multiMem}. Having an analytic framework that relates the many-body dynamics of a complex system to the reduced or \q{coarse-grained} dynamics of a low-dimensional collective variable is therefore essential for modelling and gaining a deeper understanding of the dynamics of many-body systems \cite{zwanzig2001nonequilibrium, alma_book,ayaz2022hybrid}. 
Projecting the many-body dynamics onto reaction coordinates gives rise to an effective free-energy profile \cite{Blaak_book,Cecilia_book}. The free energy profile defines states of the system, so that transitions between states can be described as free energy barrier-crossings. According to Arrhenius' law, the  barrier-crossing or mean first-passage times scale exponentially with the potential barrier height $U_0$ \cite{arrhenius1889iv}, as 

\begin{equation}
    \tau_{\text{MFP}} = \tau_* e^{U_0 \mathbin{/} k_BT} ,
    \label{eq.mfpt}
\end{equation}

\noindent where the prefactor $\tau_*$ has been predicted by transition state theory \cite{Eyring31, wigner32,wigner1938transition,laidler1983development}, Kramers theory \cite{kramers1940brownian}, reactive flux theory \cite{Keck62,anderson1973statistical,bennett1975molecular,Chandler78}, Grote-Hynes theory \cite{GH80,haynes1995reaction} and by simulations \cite{bennett1976efficient,hwang1987microscopic,bergsma1987molecular,Kohen95}. As has recently been shown \cite{Netz26}, $\tau_*$ can be related to the integral over the squared two-point correlation function of the reaction coordinate. This formulation accounts for memory and inertial effects and serves as the theoretical framework used in this work for comparison with our simulation results.\\
A systematic framework for coarse-graining is provided by the Generalized Langevin Equation (GLE) \cite{zwanzig2001nonequilibrium, peters2017reaction,Nakaj,Zwanzig, Mori}, which reduces the microscopic degrees of freedom of the many-body system to only one or a few relevant degrees of freedom through projection. Starting from the Liouville equation, the projection formalism results in a non-Markovian equation of motion that features a time-dependent friction memory function $\Gamma(t)$ and an orthogonal force $F_R(t)$, that is typically modelled as a random force.
\begin{figure*}
\centering
\includegraphics[width=0.95\textwidth]{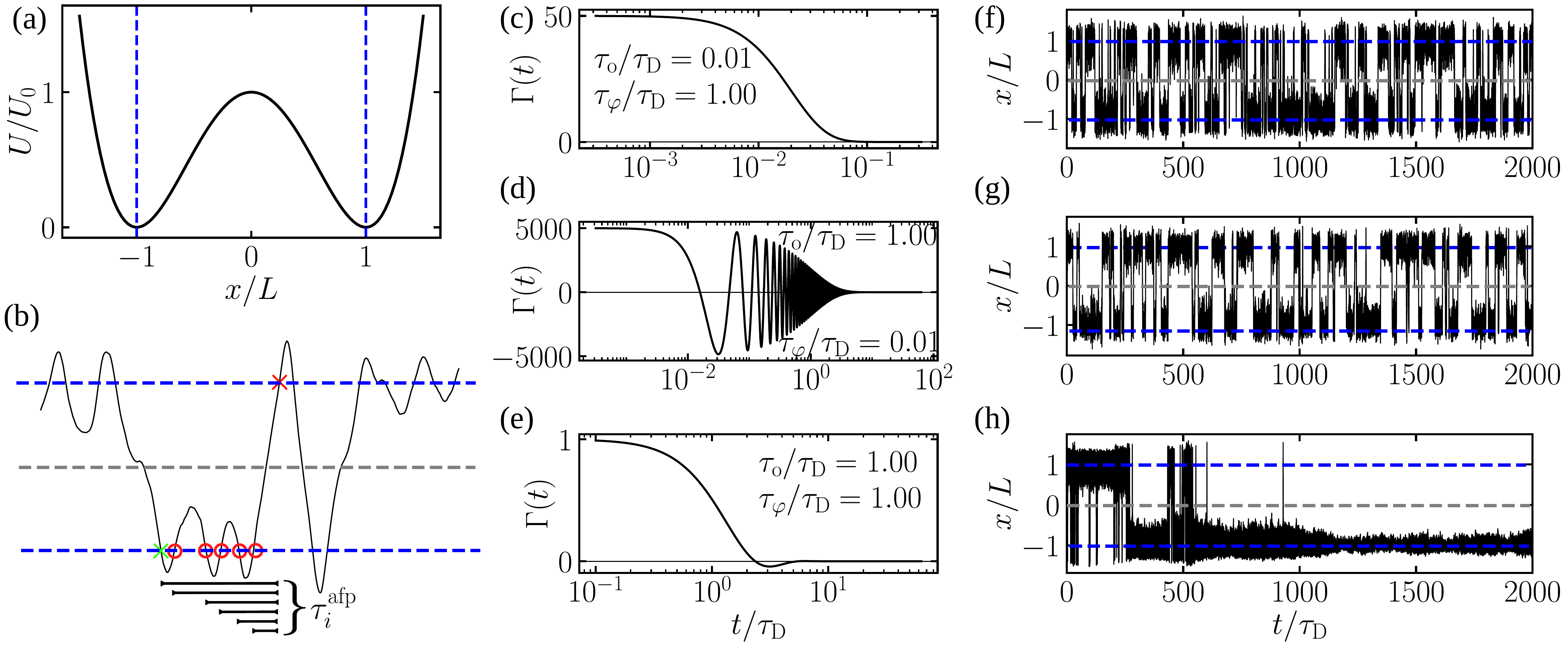} 
\caption{Barrier-crossing of a non-Markovian system with an oscillating memory kernel. (a) Symmetric double-well potential, eq.\eqref{eq.potDW}. (b) Representative simulated trajectory $x(t)$ in a double-well potential eq.(\ref{eq.potDW}) with $U_0/k_BT=3$, $\tau_\text{m}/\tau_\text{D}$ = 2, $\tau_\text{o}/\tau_\text{D}$ = 0.001 and $\tau_\varphi/\tau_\text{D}$ = 10. Blue dashed lines indicate potential minima, the grey dashed line indicates the potential barrier top. The green cross indicates the first time the trajectory reaches the left potential minimum ($x=-L$), while the red circles indicate subsequent re-crossings of the same minimum. The red cross marks the  time the trajectory reaches the opposite minimum for the first time. The time interval between the green and red crosses, together with the intervals between the red cross and the red circles, constitute the set of all first-passage events $\tau_i^{\text{afp}}$, whose average defines the mean first-passage time $\tau_{\text{MFP}}$. (c-h) Exemplary memory kernels $\Gamma(t)$ and corresponding trajectories with $\tau_{\text{m}}/\tau_{\text{D}} = 0.1$ and $U_0/k_BT=3$.  (c \& f) $\tau_{\text{o}}/\tau_{\text{D}} = 0.01, \tau_\varphi/\tau_{\text{D}} = 1.00$. 
(d \& g) $\tau_{\text{o}}/\tau_{\text{D}} = 1.00, \tau_\varphi/\tau_{\text{D}} = 0.01$. (e \& h) $\tau_{\text{o}}/\tau_{\text{D}} = 1.00, \tau_\varphi/\tau_{\text{D}} = 1.00$. }
\label{fig.one}
\end{figure*}

A wide range of systems have been successfully modelled using the GLE, such as isomerizing molecules \cite{Chandler78,bagchi1983effect,dalton2023conformational}, absorption spectra of fluids \cite{flo_line, angulo2017good}, protein folding \cite{adelman1980generalized,canales1998generalized,ayaz2021multiMem,plotkin1998non,satija2019generalized,dalton2022protein}, pair reactions of molecules \cite{flo_pai, daldrop2018butane, dalton2023conformational}, cell mobility \cite{mitterwallner2020non,klimek2024data,klimek2025intrinsic} or meteorological data \cite{Henrik_weather}. The GLE approach has also been used to analyse  non-equilibrium systems outside physics such as financial time series data \cite{Henrik_weather, hassanibesheli2020reconstructing,herrera2020tractable}. Whereas free-energy profiles can often be determined comparatively straightforwardly, extracting friction memory kernels is considerably more challenging. Recently, there has been much progress in the development of numerical tools that can directly extract friction memory kernels $\Gamma(t)$ from time series data generated from experiments or large-scale simulations \cite{straub1987calculation,horenko2007data,daldrop2018butane, hijon2010mori, carof2014two, bergsma1987molecular, vandervegt_klippenstein2021introducing, straub1986non, grogan2020data,jung2017iterative,kowalik2019memory,grogan2020data,ayaz2021multiMem,ge2024data}. The specific functional form of $\Gamma(t)$ depends on the system and can often be fitted to a sum of monotonically decaying exponential functions. However, in many instances, one also finds oscillating components \cite{ayaz2021multiMem,flo_pai,dalton2023conformational, kowalik2019memory,Darve_lee2019multi}.  

For systems with monotonically decaying friction kernels the effect of single-exponential  memory on the barrier-crossing kinetics depends on the ratio of the memory decay time $\tau$ to the diffusive timescale $\tau_\text{D}$ of the system. The barrier-crossing kinetics are accelerated for intermediate $\tau\approx \tau_\text{D} / 10$ \cite{GH80, straub1986non}, and slowed down for long $\tau>\tau_\text{D}$, showing a quadratic increase of the reaction time with $\tau$ for large $\tau$ \cite{kappler2018}, demonstrating that a simple timescale separation between environment and the barrier-crossing process does not work for long memory times \cite{dalton2025memory}.
While the influence of exponentially decaying memory on barrier-crossing kinetics is well understood from numerical studies for single-exponential \cite{GH80,straub1986non,kappler2018}, bi-exponential \cite{kappler2019non}, tri-exponential \cite{lavacchi2020trpl} kernels, as well as analytic work \cite{GH80, haynes1995reaction,Netz26}, 
comparatively little is known about the effect of oscillatory memory kernels on barrier-crossing kinetics. Apart from the general Grote–Hynes framework \cite{GH80}, dedicated studies of oscillatory memory in barrier-crossing dynamics remain scarce.

In this paper, we systematically investigate how oscillations in the friction memory kernel affect mean barrier-crossing kinetics by studying the GLE in a bistable free-energy potential using Markovian embedding simulations \cite{MarkEmb_Siegl} and analytic theory \cite{Netz26}. We find new scaling regimes of the reaction kinetics with respect to the timescales of the memory kernel: the mean first-passage times scale quartically with the memory decay time $\tau_{\text{o}}$ and the oscillation period $\tau_\varphi$ of the friction kernel for intermediate values of $\tau_{\text{o}}$ and $\tau_\varphi$, and scale linearly with $\tau_{\text{o}}$ for long values of $\tau_{\text{o}}$. Further, we show that the mean first-passage time scales as $\tau_{\text{MFP}} \sim 1/m$ for small mass $m$ and long memory decay time $\tau_{\text{o}}$ and finite $\tau_\varphi$, in stark contrast to the behaviour for non-oscillating memory kernels. Our analytical approach agrees well with our simulation results.

\begin{figure*}
\centering
\includegraphics[width=0.85\textwidth]{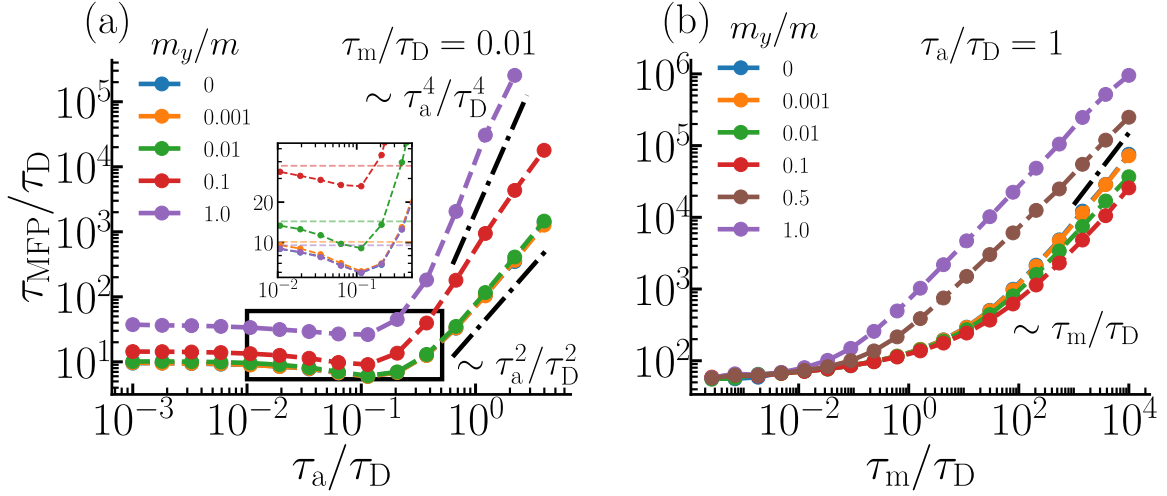} 
\caption{  Simulation results for $\tau_{\text{MFP}}$ of a GLE with a double-well potential eq.(\ref{eq.potDW}) with $U_0/k_BT=3$ and oscillating memory kernel eq.(\ref{eq.kern_osc}) for different auxiliary-to-main particle mass ratios, $m_y/m$. (a) $\tau_{\text{MFP}}/\tau_\text{D}$ as a function of $\tau_{\text{a}}/\tau_{\text{D}}$, where $\tau_{\text{a}}$ is defined in eq.\eqref{eq.taua}, with $\tau_{\text{m}}/\tau_{\text{D}}=0.01$. 
The black dash-dotted lines indicate power-law scaling. Note that the data for $m_y/m=0.01, 0.001$ and 0 are almost indistinguishable. The inset zooms into the range where memory reduces the MFPT, horizontal dashed lines indicate the Markovian limits.  (b) $\tau_{\text{MFP}}/\tau_{\text{D}}$ as a function of $\tau_{\text{m}}/\tau_{\text{D}}$ with $\tau_{\text{a}}/\tau_{\text{D}}=1$ and various ratios of auxiliary to main particle mass $m_y/m$, the black dash-dotted line indicates $\tau_{\text{MFP}}\sim \tau_{\text{m}}$.} 
\label{fig.G_osc_asymp}
\end{figure*}

\section{Results}
\label{sec.2}
\subsection*{Non-Markovian model and Markovian-embedding simulation method}
\label{sec2.1}

We consider the one-dimensional Generalized Langevin equation (GLE) for the reaction coordinate $x(t)$
\begin{equation}
    m \ddot{x}(t) = - \int^{t}_0 \Gamma(t') \dot{x}(t-t') dt' - \nabla U[x(t)] + F_R(t),
    \label{eq.gle_x}
\end{equation}

\noindent with $x$-independent mass $m$, time dependent friction kernel $\Gamma(t)$, deterministic force arising from the free energy $U[x(t)]$ and the random force $F_R(t)$ which has zero mean $ \langle F_R(t)\rangle = 0$ and second moment $ \langle F_R(t)F_R(t') \rangle = k_B T \Gamma(|t-t'|)$, thereby relating the random force to the dissipative friction kernel \cite{zwanzig2001nonequilibrium}. We consider an oscillatory-exponential memory kernel 

\begin{equation}
\begin{split}
    \Gamma_{\mathrm{osc}}(t) &=  \frac{\gamma \tau_{\text{o}}}{2} \left(\frac{1}{\tau_{\text{o}}^2}+ \frac{1}{\tau_\varphi^2} \right) e^{- t \mathbin{/} \tau_{\text{o}}} \\ &\times \left[ \cos\left(t\mathbin{/}\tau_\varphi \right) + \frac{\tau_\varphi}{\tau_{\text{o}} } \sin\left(t\mathbin{/}\tau_\varphi \right) \right] 
    \label{eq.kern_osc}        
\end{split}
\end{equation} 

\noindent with friction coefficient $\gamma = \int^{\infty}_0 \Gamma_{\mathrm{osc}}(t) dt$, exponential decay time $\tau_{\text{o}}$ and oscillation period $2 \pi \ \tau_\varphi$. The GLE eq.\eqref{eq.gle_x} for $\Gamma(t)=\Gamma_{\mathrm{osc}}(t)$ can be mapped exactly onto a system of coupled Langevin equations

\begin{subequations}
\begin{align}
    \dot{x}(t) &= v(t) \label{eq:ham_eqs_extend_x}\\
    m \dot{v}(t) &= k_y[x(t)- y(t)] - \nabla U[x(t)] \label{eq:ham_eqs_extend_v}\\
    \dot{y}(t) &= w(t) \label{eq:ham_eqs_extend_y}\\
    m_y \dot{w}(t) &= -k_y[x(t)- y(t)] - \gamma w(t) + F_y(t)\label{eq:ham_eqs_extend_w},
\end{align}
\label{eq:ham_eqs_extend}
\end{subequations}

\noindent with $v(t)$ being the velocity of the reaction coordinate $x$, and $y(t),w(t),m_y,\gamma$ being position, velocity, mass and friction of an auxiliary variable that is harmonically coupled to the reaction coordinate $x$ with coupling constant $k_y$ and is subject to a Gaussian random force $F_y(t)$ with zero mean and second moment $\langle  F_y(t)F_y(t') \rangle = 2\gamma k_B T \delta(t-t')$, see SI Sec. 1A for details of the Markovian embedding. The timescales of the kernel in eq.(\ref{eq.kern_osc}) are related to the parameters of the Langevin eqs.(\ref{eq:ham_eqs_extend}) by

\begin{align}
    \tau_{\text{o}} = \frac{2m_y}{\gamma} , && \tau_\varphi = \frac{1}{\sqrt{\frac{k_y }{m_y} - \frac{\gamma^2}{4m_y^2}}}.
    \label{eq.taus_o_phi}
\end{align}

\noindent As reference timescales, we choose 

\begin{align}
    \tau_\text{D} = \frac{\gamma L^2}{k_BT} , && \tau_{\text{m}} = \frac{m}{\gamma}.
    \label{eq.tauD_m}
\end{align}

\noindent The diffusion time $\tau_\text{D}$ is the timescale for a particle to diffuse over the length $L$ in a heat bath with thermal energy $k_BT$ and total friction $\gamma$ in the absence of a free energy profile. The inertial time $\tau_{\text{m}}$ marks the crossover between ballistic and diffusive particle motion \cite{zwanzig2001nonequilibrium}. We consider a symmetric double-well potential 

\begin{figure*}
\centering
\noindent\hspace*{-4\columnsep}%
\includegraphics[width=0.85\textwidth]{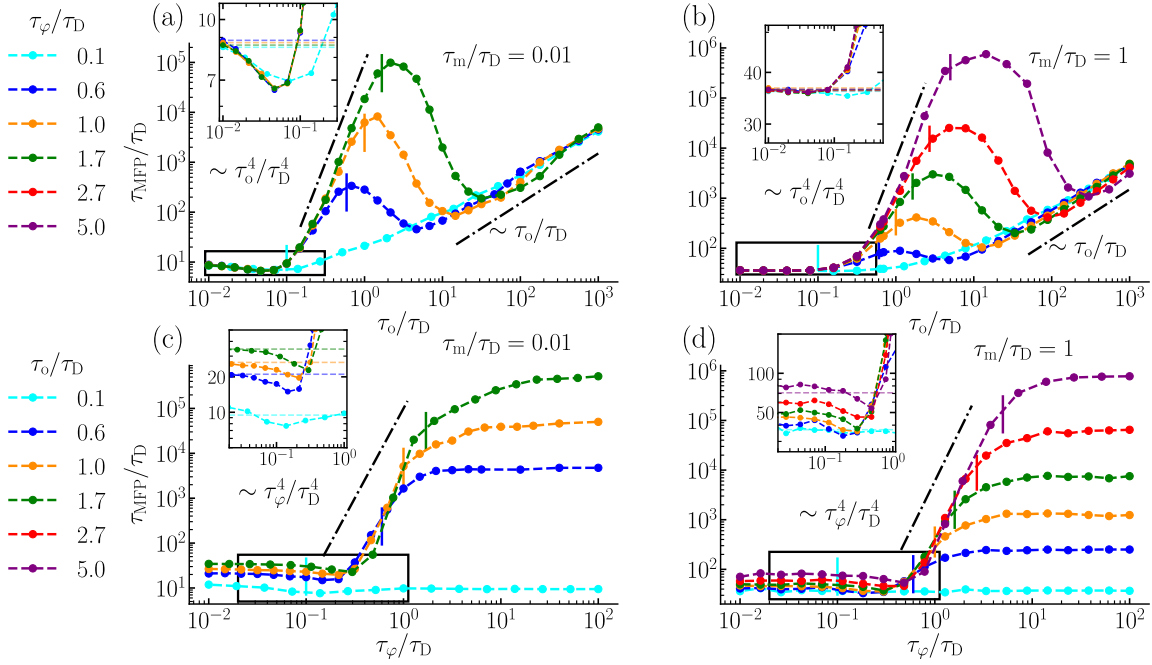} 
\caption{ 
Simulation results for the mean first-passage time $\tau_{\mathrm{MFP}}$ of a GLE system in a double-well potential, eq.(\ref{eq.potDW}), with $U_0/k_BT = 3$ and an oscillatory memory kernel, eq.(\ref{eq.kern_osc}). Results are shown as functions of the memory-kernel timescales $\tau_{\mathrm{o}}/\tau_{\mathrm{D}}$ and $\tau_\varphi/\tau_{\mathrm{D}}$.
(a,b) $\tau_{\mathrm{MFP}}/\tau_\text{D}$ as a function of $\tau_{\mathrm{o}}/\tau_{\mathrm{D}}$ for various $\tau_\varphi/\tau_{\mathrm{D}}$, at $\tau_{\mathrm{m}}/\tau_{\mathrm{D}} = 0.01$ and $\tau_{\mathrm{m}} /\tau_{\mathrm{D}}= 1$, respectively. The inset in (a) highlights the memory-induced speed-up for intermediate $\tau_{\mathrm{o}}$, dashed horizontal lines mark the Markovian limit, represented by the MFPT for the smallest value of $\tau_{\mathrm{o}}/\tau_{\mathrm{D}}$.
(c,d) $\tau_{\mathrm{MFP}}/\tau_\text{D}$ as a function of $\tau_\varphi/\tau_{\mathrm{D}}$ for various $\tau_{\mathrm{o}}/\tau_{\mathrm{D}}$, at $\tau_{\mathrm{m}}/\tau_{\mathrm{D}} = 0.01$ and $\tau_{\mathrm{m}}/\tau_{\mathrm{D}} = 1$, respectively. The insets highlight the local minima at intermediate $\tau_\varphi$, dashed horizontal lines mark the MFPT for the smallest value of $\tau_\varphi/\tau_{\mathrm{D}}$.
Black dash-dotted lines indicate the corresponding power-law scaling. Vertical bars mark $\tau_{\mathrm{o}} = \tau_\varphi$. 
}
\label{fig.OSC_MFPT_SCALING_COMP_THEORY}
\end{figure*}

\begin{equation}
U(x)= U_0\left[\left(\frac{x}{L}\right)^2 - 1\right]^2,    
    \label{eq.potDW}
\end{equation}

\noindent with barrier height $U_0$ and well separation $2L$, see Figure \ref{fig.one} a. We quantify the barrier-crossing time by the mean first-passage time $\tau_{\text{MFP}}$ (MFPT). The $\tau_{\text{MFP}}$ is obtained from the mean of all first-passage events between $x=-L$ and $x=L$ in a long continuous simulation trajectory, see Figure \ref{fig.one} b; numerical details are provided in SI Sec. 1B–C. This definition of $\tau_{\text{MFP}}$ corresponds to the reaction or escape time \cite{Recross_Ben}, see SI Sec. 2 for details. We simulate the reaction kinetics by integrating eqs.(\ref{eq:ham_eqs_extend}) and analyse 
the barrier-crossing kinetics as a function of the four characteristic timescales $\{\tau_\text{D},\tau_\text{m},\tau_{\text{o}},\tau_\varphi\}$ of the GLE.

In Figure \ref{fig.one} c--h, we show oscillatory memory kernels for different memory times $\{\tau_\text{o}/\tau_\text{D},\tau_\varphi/\tau_\text{D} \}$ and  the corresponding trajectories for $\tau_\text{m}/\tau_\text{D}=0.1$ and $U_0/k_BT=3$. We observe frequent barrier-crossing events in Figure \ref{fig.one} f and g, but in Figure \ref{fig.one} h, when both the exponential decay time $\tau_\text{o}$ and the oscillation period $\tau_\varphi$ are set equal to the diffusion time, the trajectory exhibits extended phases of rapid state re-crossings \cite{Recross_Ben} and long phases without transitions, typical for GLE systems in the regime of memory-induced slowdown \cite{kappler2018}. These observations suggest that for GLE systems with oscillatory memory, the slowdown of the reaction kinetics is strongest when both $\tau_\text{o}$ and $\tau_\varphi$ are long compared to $\tau_\text{D}$.

\subsection*{Limit of monotonically decaying exponential memory}
\label{sec2.2}

We start with a discussion of the
oscillatory-exponential memory in terms of the parameters of the  Markovian embedding eqs.\eqref{eq:ham_eqs_extend}, because in this formulation the monotonically decaying exponential memory limit is straightforwardly obtained by considering $m_y \to 0$. In this limit, the oscillatory-exponential kernel eq.(\ref{eq.kern_osc}) converges to a purely exponentially decaying function 

\begin{align}
 \lim_{m_y\to 0} \Gamma_{\text{osc}}(t) & \approx     k_y e^{- t \; k_y \mathbin{/} \gamma} = \frac{\gamma}{\tau_{\text{a}}} e^{- t \; \mathbin{/} \tau_{\text{a}}}
    \label{eq.kernel_limit_osc_exp}
\end{align}

\noindent with the asymptotic decay time given by

\begin{equation}
    \tau_{\text{a}} = \gamma/k_y,
    \label{eq.taua}
\end{equation}
\noindent see Appendix \ref{sec.lim_osc-exp} for details. In Figure \ref{fig.G_osc_asymp} a, we show $\tau_{\text{MFP}}/\tau_\text{D}$ as a function of $\tau_{\text{a}}/\tau_\text{D}$ for various ratios $m_y/m$. For small $m_y/m$, we recover the previously studied single-exponential kernel scenario \cite{kappler2018}: for small decay times $\tau_\text{a} \ll \tau_\text{D}$,  $\tau_{\text{MFP}}$ converges to the Markovian limit; for intermediate $\tau_{\text{a}}/\tau_\text{D} \approx 0.1$, the barrier-crossing dynamics is accelerated, as reflected by a decrease in $\tau_{\text{MFP}}$, as shown in the inset of Figure \ref{fig.G_osc_asymp} a. For long $\tau_{\text{a}}\gg\tau_\text{D}$, the barrier-crossing kinetics slows down, as reflected by a quadratic increase of $\tau_{\text{MFP}}$ with $\tau_{\text{a}}$. Increasing the ratio $m_y/m$ turns the quadratic into  a quartic scaling, see Figure \ref{fig.G_osc_asymp} a. We also show $\tau_{\text{MFP}}$ for $m_y / m = 0$, which corresponds to a purely exponential kernel, exhibiting perfect agreement with the results for $m_y / m = 0.001$. In Figure \ref{fig.G_osc_asymp} b, we show $\tau_{\text{MFP}}/\tau_\text{D}$ as a function of $\tau_{\text{m}}/\tau_\text{D}$ for various ratios $m_y/m$, where we observe the Kramers turnover \cite{kramers1940brownian,zwanzig2001nonequilibrium}: $\tau_{\text{MFP}} \sim 1/\gamma$ for low friction and $\tau_{\text{MFP}} \sim \gamma$ for high friction. In terms of our dimensionless variables, this translates to $\tau_{\text{MFP}}/\tau_{\text{D}} \sim \tau_\text{m}/\tau_\text{D}$ for large $\tau_\text{m}/\tau_\text{D}$ (small friction) and $\tau_{\text{MFP}}/\tau_{\text{D}} \sim \text{const.}$ for small $\tau_\text{m}/\tau_\text{D}$ (high friction). For large values of $\tau_{\text{m}}/\tau_{\text{D}}$ and finite $m_y/m$, $\tau_{\text{MFP}}$ exhibits an additional turnover arising from the inertia of the auxiliary system. In this regime, $\tau_{\text{MFP}}$ initially decreases as the mass ratio $m_y / m$ is reduced from $1.0$ to $0.1$, but subsequently increases as $m_y / m$ is further reduced from $0.01$ to $0$.

\subsection*{$\tau_{\mathrm{MFP}}$ for oscillatory-exponential memory kernel}
\label{sec2.3}

\begin{figure*}
\centering
\includegraphics[width=0.99\textwidth]{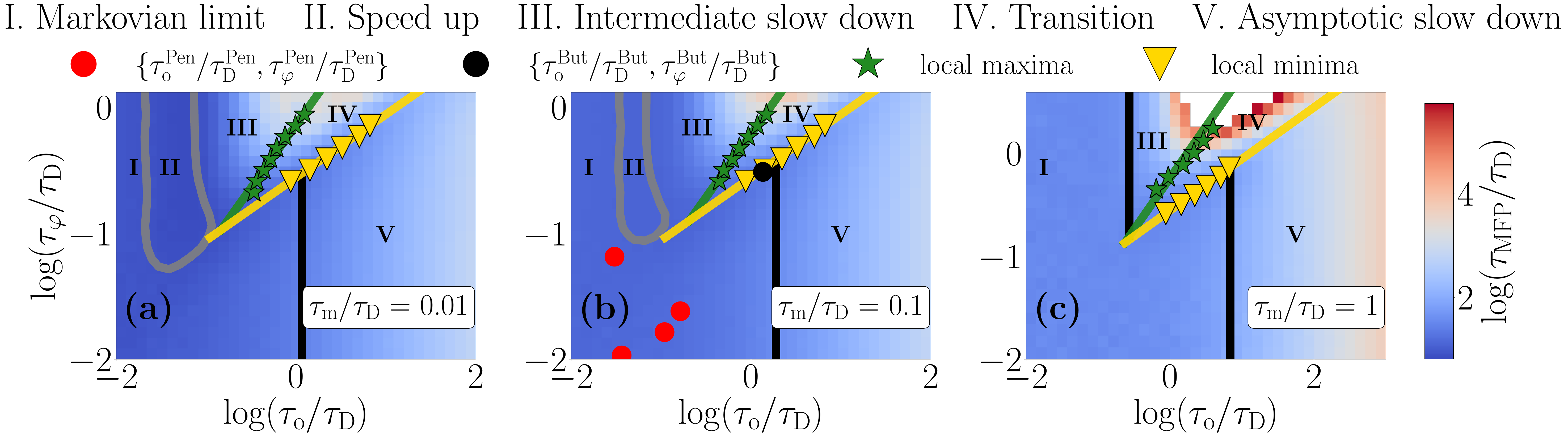} 
\caption{
Scaling diagrams of $\tau_{\mathrm{MFP}}/\tau_{\mathrm{D}}$ as a function of $\tau_{\mathrm{o}}/\tau_{\mathrm{D}}$ and $\tau_\varphi/\tau_{\mathrm{D}}$ for different inertial times $\tau_{\mathrm{m}}/\tau_{\mathrm{D}}$. Five distinct regimes of the reaction dynamics are identified: (I) Markovian limit, (II) memory speed-up regime, (III) intermediate slowdown regime with $\tau_{\mathrm{MFP}} \sim \tau_{\mathrm{o}}^4$, (IV) transition regime with $\tau_{\mathrm{MFP}} \sim \tau_{\varphi}^4$, and (V) asymptotic slowdown regime with $\tau_{\mathrm{MFP}} \sim \tau_{\mathrm{o}}$. The grey contour, shown after smoothing for visual clarity, encloses the memory speed-up regime, where $\tau_{\mathrm{MFP}}$ is reduced by more than $5\%$ relative to the Markovian limit. Green stars and cyan triangles denote local maxima and minima of $\tau_{\mathrm{MFP}}$ as a function of $\tau_{\mathrm{o}}$ at fixed $\tau_\varphi$, respectively. The corresponding fitted scaling relations are $\tau_\varphi \sim \tau_{\mathrm{o}}$ (green line) and $\tau_\varphi \sim \sqrt{\tau_{\mathrm{o}}}$ (yellow line). Black solid lines indicate boundaries between the Markovian and slowdown regimes.
(a) $\tau_{\mathrm{m}}/\tau_{\mathrm{D}} = 0.01$.
(b) $\tau_{\mathrm{m}}/\tau_{\mathrm{D}} = 0.1$; red circles indicate oscillating memory components obtained from MD simulations of pentane isomerization in water \cite{Henrik_multi}, while the black circle indicates the oscillating memory component obtained from MD simulations of butane isomerization in water \cite{dalton2023conformational}.
(c) $\tau_{\mathrm{m}}/\tau_{\mathrm{D}} = 1$.
}
\label{fig.phase}
\end{figure*}

In the preceding section, we presented the barrier-crossing times in terms of the parameters appearing in the Markovian embedding eqs.\eqref{eq:ham_eqs_extend}. In experiments and simulations, however, the extracted memory kernel is naturally parameterized by eq.\eqref{eq.kern_osc}. Therefore, from now on, we discuss $\tau_{\mathrm{MFP}}/\tau_{\mathrm{D}}$ as a function of the timescales $\tau_{\mathrm{o}}/\tau_{\mathrm{D}}$ and $\tau_\varphi/\tau_{\mathrm{D}}$ characterizing the oscillating friction kernel $\Gamma_{\mathrm{osc}}(t)$. Infact, the scaling of $\tau_{\mathrm{MFP}}$ as a function of $\tau_{\mathrm{o}}/\tau_{\mathrm{D}}$ and $\tau_\varphi/\tau_{\mathrm{D}}$ differs qualitatively from the results shown in Figure \ref{fig.G_osc_asymp}, despite the simple relations in eq.\eqref{eq.taus_o_phi}, as will be discussed below. 

In Figure \ref{fig.OSC_MFPT_SCALING_COMP_THEORY} a, we show $\tau_{\text{MFP}}/\tau_{\text{D}}$ for fixed $\tau_{\text{m}}/\tau_{\text{D}}=0.01$ as a function of $\tau_{\text{o}}/\tau_{\text{D}}$ for a few different values of $\tau_\varphi/\tau_{\text{D}}$; for $\tau_{\text{o}}/\tau_{\text{D}}\rightarrow 0$, we recover the Markovian limit. The decrease of $\tau_{\text{MFP}}$ between $0.01<\tau_{\text{o}}/\tau_{\text{D}}<0.1$ reflects the memory-induced speed-up of the barrier-crossing dynamics, where $\tau_{\text{MFP}}$ drops below the Markovian limit (see inset in Figure \ref{fig.OSC_MFPT_SCALING_COMP_THEORY} a). For intermediate values $0.1<\tau_{\text{o}}/\tau_{\text{D}}<1$, $\tau_{\text{MFP}}$ increases as $\tau_\text{o}^4/\tau_\text{D}^4$, followed by an intermediate decline of $\tau_\text{MFP}$ and an asymptotic linear increase $\tau_{\text{MFP}} \sim \tau_\text{o}$ for large $\tau_\text{o}/\tau_\text{D}$. Thus, an oscillatory-exponential memory kernel induces both accelerating and slowing down effects and gives rise to two minima and two distinct power laws when plotting $\tau_{\text{MFP}}$ as a function of $\tau_{\text{o}}$, see Figures \ref{fig.OSC_MFPT_SCALING_COMP_THEORY} a-b. The dependence of $\tau_{\text{MFP}}$ on $\tau_\text{o}$ changes significantly with the value of $\tau_\varphi$: for $\tau_\varphi \ll \tau_\text{D}$, the scaling $\sim \tau_\text{o}^4$ is not present, and the MFPT rises monotonically with $\tau_\text{o}$ for $\tau_{\text{o}}/\tau_{\text{D}}>0.1$ (see cyan curves in Figures \ref{fig.OSC_MFPT_SCALING_COMP_THEORY} a-b). With increasing $\tau_\varphi/\tau_{\text{D}}$, the range over which the scaling $\tau_{\text{MFP}}\sim\tau_{\text{o}}^4$ is observed increases. When comparing the results for $\tau_{\mathrm{m}}/\tau_{\mathrm{D}}=0.01$ and $\tau_{\mathrm{m}}/\tau_{\mathrm{D}}=1$ in Figures ~\ref{fig.OSC_MFPT_SCALING_COMP_THEORY} a-b, we observe that the memory-induced speed-up, present in the range $0.01<\tau_{\mathrm{o}}/\tau_{\mathrm{D}}<0.1$ for $\tau_{\mathrm{m}}/\tau_{\mathrm{D}}=0.01$, is absent for $\tau_{\mathrm{m}}/\tau_{\mathrm{D}}=1$. We also observe an acceleration of the reaction dynamics with increasing the inertial timescale for fixed $\tau_\varphi / \tau_\text{D}$. This behaviour is characteristic of oscillatory memory effects and will be discussed in more detail below.

In Figure \ref{fig.OSC_MFPT_SCALING_COMP_THEORY} c-d, we show $\tau_{\text{MFP}}/\tau_{\mathrm{D}}$ as a function of $\tau_\varphi/\tau_{\text{D}}$ for two different values of $\tau_{\text{m}}/\tau_{\text{D}}$ and a few different values of $\tau_{\text{o}}/\tau_{\text{D}}$. The MFPT approaches constant plateaus for $\tau_\varphi\rightarrow0$, reduces with increasing $\tau_\varphi$ for $\tau_\varphi<\tau_{\text{D}}$ (see inset in Figure \ref{fig.OSC_MFPT_SCALING_COMP_THEORY} d) and subsequently increases monotonically, scaling as the fourth power of $\tau_\varphi/\tau_{\text{D}}$, before reaching a plateau as $\tau_\varphi\rightarrow\infty$. At small $\tau_\text{o}$, the $\sim\tau_\varphi^4$ scaling is absent (see cyan curves in Figure \ref{fig.OSC_MFPT_SCALING_COMP_THEORY} c-d), while the intermediate regime with the quartic scaling $\tau_{\text{MFP}}\sim\tau_\varphi^4$ widens as $\tau_\text{o}$ increases. Thus, the effect of the oscillation period $\tau_\varphi$ of an oscillatory-exponential memory kernel is most significant for intermediate $\tau_\varphi\approx\tau_\text{D}$ and is qualitatively the same for $\tau_\text{m}/\tau_\text{D}=0.01$ and $\tau_\text{m}/\tau_\text{D}=1$.

Figure~\ref{fig.phase} summarizes the scaling behavior of $\tau_{\mathrm{MFP}}/\tau_\text{D}$ as a function of the memory timescales $\tau_{\mathrm{o}}/\tau_{\mathrm{D}}$ and $\tau_{\varphi}/\tau_{\mathrm{D}}$ for three different values of $\tau_{\mathrm{m}}/\tau_{\mathrm{D}}$. Five distinct regimes of the reaction dynamics are identified: (I) the Markovian limit at small $\tau_{\mathrm{o}}$, (II) the memory speed-up regime, where the MFPT is reduced by more than 5\% compared to the Markovian limit and which is enclosed by grey solid lines, (III) the intermediate slowdown regime, characterized by $\tau_{\mathrm{MFP}} \sim \tau_{\mathrm{o}}^4$ and $\tau_{\mathrm{MFP}} \sim \tau_\varphi^0$, (IV) the transition regime, characterized by $\tau_{\mathrm{MFP}} \sim \tau_{\varphi}^4$ and a decrease of $\tau_{\mathrm{MFP}}$ with $\tau_{\mathrm{o}}$ and (V) the asymptotic slowdown, characterized by $\tau_{\mathrm{MFP}} \sim \tau_{\mathrm{o}}$ for long $\tau_{\mathrm{o}}$ and $\tau_{\mathrm{MFP}}  \sim \tau^0_{\varphi}$. The boundaries between these regimes are determined from the characteristic extrema and scaling crossovers of the MFPT, as discussed below. The green stars mark maxima of $\tau_{\mathrm{MFP}}$ as a function of $\tau_{\mathrm{o}}$ for fixed $\tau_\varphi$ (see Figures \ref{fig.OSC_MFPT_SCALING_COMP_THEORY} a-b). These maxima mark the transition between regimes (III) and (IV) and are well described by the scaling $ \tau_\varphi \sim \tau_{\mathrm{o}}$ (green solid line). The yellow triangles mark minima of  $\tau_{\mathrm{MFP}}$ as a function of $\tau_{\mathrm{o}}$ for fixed $\tau_\varphi$ (see Figures \ref{fig.OSC_MFPT_SCALING_COMP_THEORY} a-b), indicating the transition between regimes (IV) and (V) as well as between regimes (IV) and (I). This transition is well described by a scaling $\tau_\varphi \sim \sqrt{\tau_{\mathrm{o}}}$ (yellow solid line). The vertical black solid lines separating regimes (I) and (V) are obtained from the mean intersection point of linear fits to the small-$\tau_{\mathrm{o}}$ plateau and the large-$\tau_{\mathrm{o}}$ linear regime, as seen in Figures \ref{fig.OSC_MFPT_SCALING_COMP_THEORY} a-b. The vertical black solid line separating regimes (I) and (III) in Figure \ref{fig.phase} c is obtained from the mean  intersection points of fits to the plateau regime at small $\tau_{\mathrm{o}}$ and to the superlinear scaling of $\tau_{\mathrm{MFP}}$ at intermediate $\tau_{\mathrm{o}}$, as seen in Figure \ref{fig.OSC_MFPT_SCALING_COMP_THEORY} b. The black circle in Figure \ref{fig.phase} b represents the oscillatory memory timescale values extracted from MD simulations 
of butane isomerization in water \cite{dalton2023conformational}. Here, the parameters are located at the boundary between regimes (IV) and (V), which demonstrates that the butane isomerization exhibits complex barrier-crossing dynamics as dictated of its memory kernel parameters. Red circles in Figure \ref{fig.phase} b mark timescale values of four of five oscillating components obtained from MD simulation of  pentane isomerization in water \cite{Henrik_multi} (the fifth component falls outside the range shown in the figure), demonstrating that the pentane isomerization dynamics is completely different from butane and governed by Markovian theory. 

\begin{figure*}
\centering
\includegraphics[width=0.99\textwidth]{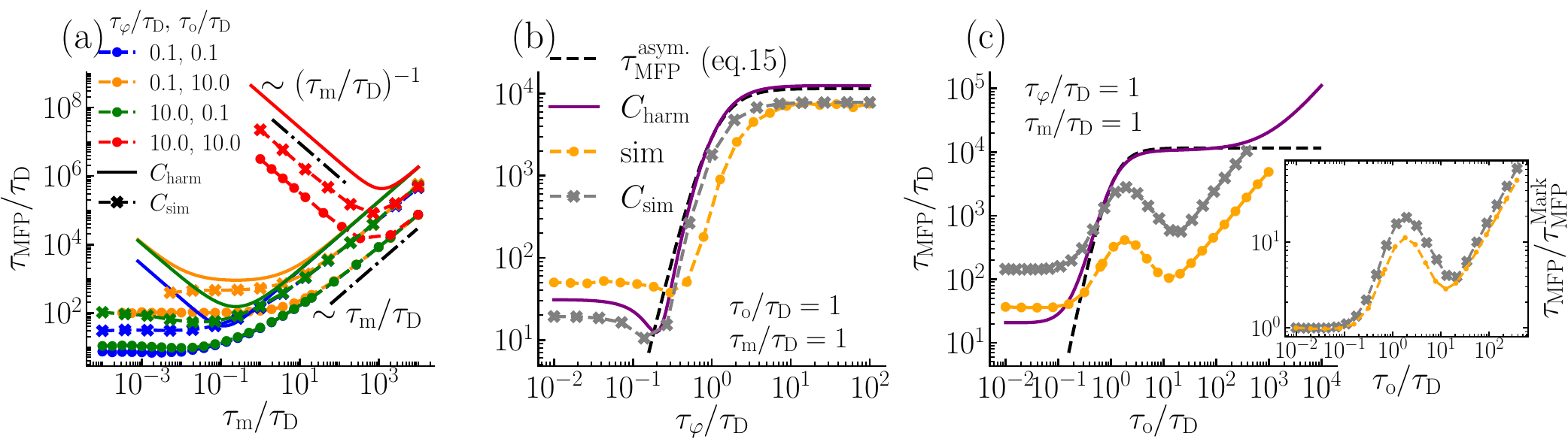} 
\caption{
Simulation results for the mean first-passage time $\tau_{\mathrm{MFP}}$ of the GLE  in a double-well potential, eq.(\ref{eq.potDW}), with $U_0/k_BT = 3$ and an oscillatory memory kernel, eq.(\ref{eq.kern_osc}), together with theoretical predictions.
(a) $\tau_{\text{MFP}}$ as a function of $\tau_{\text{m}}/\tau_{\text{D}}$, for different timescales of oscillatory memory kernel $\tau_{\text{o}}/\tau_{\text{D}}$ and $\tau_\varphi/\tau_{\text{D}}$. Black dashed-dotted lines indicate scaling behaviour in powers of $\tau_{\text{m}}/\tau_{\text{D}}$.
In all panels, simulation results (circles and dashed lines) are compared to theoretical predictions of $\tau_{\mathrm{MFP}}$ according to eqs.~(\ref{eq.mfpt},\ref{eq.tau_rel}), using either the analytic correlation function $C_{\mathrm{harm}}$ (solid lines) or the simulated correlation functions $C_{\mathrm{sim}}$ (dashed lines with cross markers). (b)  $\tau_{\mathrm{MFP}}$ as a function of $\tau_\varphi/\tau_{\mathrm{D}}$ for $\tau_{\mathrm{o}}/\tau_{\mathrm{D}} = 1$. (c) $\tau_{\mathrm{MFP}}$ as a function of  $\tau_{\mathrm{o}}/\tau_{\mathrm{D}}$ for $\tau_\varphi/\tau_{\mathrm{D}} = 1$. In both panels, $\tau_{\mathrm{m}}/\tau_{\mathrm{D}} = 1.$
The black dashed lines in panel (b) and (c) show the asymptotic expansion of $\tau_{\mathrm{MFP}}$ according to eq.\eqref{eq.mfpt_scaling}. The inset in (c) shows $\tau_{\mathrm{MFP}}/\tau_{\mathrm{MFP}}^\text{Mark}$, where $\tau_{\mathrm{MFP}}^\text{Mark}$ denotes the MFPT for with the smallest value of $\tau_{\mathrm{o}}$, approximating the Markovian limit.
} 
\label{fig.OSC_tau_m}
\end{figure*}

In Figure \ref{fig.OSC_tau_m} a, we show simulation results of $\tau_{\text{MFP}}/\tau_\text{D}$ as a function of
$\tau_{\text{m}}/\tau_{\text{D}}$ for several values of
$\tau_{\text{o}}/\tau_{\text{D}}$ and $\tau_\varphi/\tau_{\text{D}}$ (circles connected by broken lines) and a comparison to theoretical predictions (solid and dashed curves with cross markers), which are derived in the following section. We observe that the Markovian limiting behavior, characterized by the Kramers turnover from the overdamped scaling
$\tau_{\text{MFP}}\sim\gamma$ to the underdamped scaling
$\tau_{\text{MFP}}\sim1/\gamma$ (in our variables, this corresponds to  $\tau_{\text{MFP}}/\tau_{\text{D}} \sim  \text{const}.$ and $\tau_{\text{MFP}}/\tau_{\text{D}} \sim \tau_\text{m}/\tau_\text{D}$, respectively), is in the simulation data recovered only when at least one of the ratios
$\tau_{\text{o}}/\tau_{\text{D}}$ or $\tau_\varphi/\tau_{\text{D}}$ is smaller than unity. 
If both $\tau_{\text{o}} , \tau_\varphi \gtrapprox \tau_{\text{D}}$, the simulation barrier-crossing dynamics (red circles) exhibit a memory-induced slowdown when reducing the inertial time, with $\tau_{\text{MFP}}$ converging to $\tau_{\text{MFP}} \sim 1/m $, corresponding to $\tau_{\text{MFP}}/\tau_{\text{D}} \sim (\tau_\text{m}/\tau_{\text{D}} )^{-1} $. This behavior emerges before reaching the overdamped limit $\tau_{\text{m}}/\tau_{\text{D}}<1$. This deviation of  $\tau_{\text{MFP}}$ from the Markovian scaling as a function of $\tau_\text{m}$ marks a novel feature absent for system with non-oscillating exponential memory kernel components. Note that the $\tau_{\text{MFP}} \sim 1/m $ scaling for small $\tau_\text{m}/\tau_{\text{D}}$ is not present when $\tau_{\text{MFP}}$ is plotted as a function of $\tau_\text{m}/\tau_{\text{D}}$ for fixed $\tau_\text{a}/\tau_\text{D}$ and $m_y/m$ due to the non-linear parameter rescaling eq.\eqref{eq.taus_o_phi}, see Figure \ref{fig.G_osc_asymp} b. This can be understood from the fact that for constant $\tau_\text{o}/\tau_{\text{D}}$ and $\tau_\varphi/\tau_{\text{D}}$, the value of $m_y/m$ diverges as $\tau_\text{m}/\tau_{\text{D}}\rightarrow 0$, as follows from 
\begin{equation}
    \frac{m_y}{m} = \frac{\tau_\text{o}/\tau_{\text{D}} }{ 2 \tau_\text{m}/\tau_{\text{D}} }.
    \label{eq.my_overm_tau_o}
\end{equation}

\noindent So the limit $\tau_\text{m}/\tau_{\text{D}}\rightarrow 0$ while $\tau_\text{o}/\tau_{\text{D}} = \text{const}.$ does not correspond to the limit $\tau_\text{m}/\tau_{\text{D}}\rightarrow 0$ while $m_y/m = \text{const}.$, which is considered in Figure \ref{fig.G_osc_asymp} b.

\subsection*{Comparison with analytical description of non-Markovian barrier-crossing dynamics}
\label{sec2.4}

For a Markovian overdamped (i.e. mass-less) system, the barrier-crossing time $\tau_{\text{MFP}}$ is given by 

\begin{equation}
    \tau_{\text{MFP}} = \frac{1}{D_0} \int^{x_\text{f}}_{x_\text{s}} dx e^{\beta U(x) } \int^{x}_{x_\text{ref}} dx' e^{-\beta U(x') },
    \label{eq.mpft_FP}
\end{equation}

\noindent with $x_\text{s},x_\text{f}$ being the starting and final positions, $x_\text{ref}<x_s$ being the position of a reflecting boundary condition imposed on the reactive particle flux \cite{zwanzig2001nonequilibrium}, $\beta= 1/k_BT$ and diffusivity $D_0$. For a general non-Markovian inertial system, it has been recently shown \cite{Netz26} that within a cumulant expansion and using a harmonic approximation, the diffusivity in eq.\eqref{eq.mpft_FP} can be written
as

\begin{equation}
    \frac{1}{D_0} = \frac{\tau_\text{rel}}{C_0},
        \label{eq.diff}
\end{equation}

\noindent with $C_0=C(t=0)=\langle x^2\rangle$ being the mean square of the reaction coordinate. The relaxation time $\tau_\text{rel}$ is given by the integral of the squared two-point correlation function $C(t) = \langle x(t) x(0)\rangle $ 

\begin{equation}
    \tau_{\text{rel}} = \frac{2}{C^2_0} \int_{0}^{\infty}  dt  C^2(t).
    \label{eq.tau_rel}
\end{equation}

\noindent The double integral in Eq.~\eqref{eq.mpft_FP} can be evaluated within a saddle-point approximation for $x_s$ to the left and $x_f$ to the right of the barrier, leading to the Arrhenius-type expression 

\begin{equation}
    \tau_{\text{MFP}} = 2\pi \sqrt{\frac{K_\text{min}}{K_\text{bar}}} \tau_\text{rel}  e^{\beta U_0},
    \label{eq.mfpt_rel_theta}
\end{equation}

\noindent for the double-well potential defined in eq.~\eqref{eq.potDW}. Here, $K_\text{min}=8U_0/L^2$ and $K_\text{bar}=4U_0/L^2$ denote the curvature magnitudes of the potential at the minimum and at the barrier, respectively, see Sec.~3 of the SI for further details. Eq. \eqref{eq.mfpt_rel_theta} shows that the pre-exponential factor $\tau_*$, defined in eq.\eqref{eq.mfpt}, is given by the product of the relaxation time $\tau_\text{rel} $ and a factor coming from the double integral over the free energy profile $U(x)$ in eq.\eqref{eq.mpft_FP}. We evaluate the integral in eq.\eqref{eq.tau_rel} for the correlation function $C_\text{harm}(t)$ obtained for a harmonic potential by residual calculus, see SI Sec. 4 for details. This reproduces the scaling behavior of the MFPT as functions of  $\tau_{\text{o}}$ and $\tau_\varphi$, see solid purple curves in Figures \ref{fig.OSC_tau_m} b-c.
We note that when employing $C_\text{harm}(t)$ in eq.\eqref{eq.tau_rel}, the resulting $\tau_{\text{MFP}}/\tau_\mathrm{D}$ as a function of $\tau_\text{o}/\tau_\text{D}$ does not reproduce the minimum separating the $\tau_\text{MFP}\sim\tau_\text{o}^4$ and $\tau_\text{MFP}\sim\tau_\text{o}$ scaling regimes, see purple curve in Figure \ref{fig.OSC_tau_m} c. In contrast, when the correlation function $C_\text{sim}(t)$, extracted from simulation, is used in eq.(\ref{eq.tau_rel}), the resulting $\tau_{\text{MFP}}$ captures the non-monotonic dependence on $\tau_\text{o}/\tau_\text{D}$ in the transition regime (grey crosses connected by broken line in Figure \ref{fig.OSC_tau_m} c). We take the MFPT obtained for the smallest simulated value of $\tau_\text{o}$ as the Markovian limit reference value, denoted by $\tau_\text{MFP}^\text{Mark}$. Dividing both the simulated MFPTs and the MFPT inferred from $C_\text{sim}(t)$ by the respective values of $\tau_\text{MFP}^\text{Mark}$, we find good agreement between simulation results and the theoretical prediction (see inset in Figure \ref{fig.OSC_tau_m} c). 
When comparing the simulated MFPT as a function of $\tau_{\mathrm{m}}/\tau_{\mathrm{D}}$ with the theoretical predictions in Figure~\ref{fig.OSC_tau_m} a, two observations can be made. First, the prediction based on $C_{\mathrm{harm}}(t)$ (solid lines) correctly captures the $\tau_{\mathrm{MFP}}\sim1/\tau_{\mathrm{m}}$ divergence for $\tau_{\mathrm{o}},\tau_\varphi\gtrsim\tau_\mathrm{D}$, but predicts this divergence for all values of $\tau_{\mathrm{o}}/\tau_{\mathrm{D}}$ and $\tau_\varphi/\tau_{\mathrm{D}}$, in disagreement with the simulated plateau observed whenever at least one of the two memory timescales is smaller than $\tau_{\mathrm{D}}$. Second, replacing $C_{\mathrm{harm}}(t)$ by the simulated correlation function $C_{\mathrm{sim}}(t)$ recovers both the $\tau_{\mathrm{MFP}}\sim1/\tau_{\mathrm{m}}$ scaling for $\tau_{\mathrm{o}},\tau_\varphi\gtrsim\tau_{\mathrm{D}}$ and the plateau for the case when at least one of the values of $\tau_{\mathrm{o}}/\tau_{\mathrm{D}}$ or $\tau_\varphi/\tau_{\mathrm{D}}$ is small.

The discrepancy between eq.\eqref{eq.mfpt_rel_theta} and simulated MFPTs is also observed when the trajectories are generated in a harmonic potential, see SI Sec. 5 for details. We conclude that the deviations do not stem from the harmonic approximation of the potential in calculating eq.\eqref{eq.tau_rel}, but rather are caused by inherent approximations of the theory, which involve the saddle-point approximation of eq.\eqref{eq.mpft_FP} and the cumulant expansion of the time-dependent diffusivity leading to eq.\eqref{eq.tau_rel} \cite{Netz26}.

Analysing the asymptotic behaviour of $ \lim_{m\rightarrow0}  \tau_{\text{MFP}}$ based on eq.\eqref{eq.tau_rel} with $C_\text{harm}(t)$, we find

\begin{equation}
    \lim_{m\rightarrow0}  \tau_{\text{MFP}} = e^{U_0/k_B T} \frac{1}{m}  \frac{K^2}{\gamma } \frac{\tau_\text{o}^4 \tau_\varphi^4}{(\tau_\text{o}^2 + \tau_\varphi^2)^2}   + \mathcal{O}\left(m^0\right),
    \label{eq.mfpt_scaling}
\end{equation}

\noindent see SI Sec. 4 for the derivation. Eq.(\ref{eq.mfpt_scaling}) predicts the $\tau_\text{MFP}\sim\tau_\text{o}^4$ and the $\tau_\text{MFP}\sim\tau_\varphi^4$ scaling for $\tau_\text{o}<\tau_\varphi$ and $\tau_\text{o}>\tau_\varphi$, respectively (see black dashed lines in Figure \ref{fig.OSC_tau_m} b-c). It also correctly predicts the $1/m$ scaling of the barrier-crossing time for small mass. In the limit of small auxiliary mass $m_y$, the leading 
$ \mathcal{O}\left(1/m\right)$ contribution in Eq.~(\ref{eq.mfpt_scaling}) vanishes, giving rise to $ \mathcal{O}\left(m^0\right)$ scaling, in agreement with the previously derived result for systems with purely exponential memory \cite{Netz26}.


\section{Conclusion}
\label{sec3}
We analyse the effect of an oscillating memory kernel on barrier-crossing kinetics in a symmetric double-well potential using simulations and analytical methods. We identify novel scaling behaviour of the MFPT $\tau_{\text{MFP}}$ with respect to the timescales $\tau_{\text{o}}$ and $\tau_\varphi$ of the oscillating memory.
We construct a scaling diagram that features regimes in which $\tau_{\text{MFP}}$ exhibits different power-law scalings as a function of $\tau_{\text{o}}$ and $\tau_\varphi$. For the case of butane isomerisation in water, the memory kernel parameters are located at the boundary between two scaling regimes, indicating that small changes in system parameters (temperature, presence of cosolutes) will change the barrier-crossing characteristics in an intricate fashion \cite{dalton2023conformational}. Additionally, we find that the barrier-crossing time scales as $\tau_{\text{MFP}}\sim 1/m$ for small mass of the reaction coordinate, in contrast to the case of Markovian or non-Markovian systems with purely exponential kernels, which originates from the inertia associated with oscillatory memory and, within the exact Markovian embedding eqs.\eqref{eq:ham_eqs_extend}, is represented by the finite mass of the auxiliary coordinate. Our analytical approach, which relates the prefactor of the Arrhenius expression for the barrier-crossing time to the integral of the squared correlation function \cite{Netz26},
nicely reproduces the scaling behaviour of $\tau_{\text{MFP}}$ from the simulations. As oscillatory kernel components are widely encountered in the analysis of time-series data within the GLE framework \cite{ayaz2021multiMem,flo_pai,dalton2023conformational, kowalik2019memory}, our analysis of GLE reaction dynamics with competing timescales contributes to the development of a generalized theory of non-Markovian reaction dynamics.

An important direction for future work is the extension of the present analysis to more realistic memory kernels composed of multiple components. In many systems, extracted memory kernels contain both a monotonically decaying contribution and one or several oscillatory components \cite{flo_line, flo_pai, dalton2023conformational}. Understanding how these competing contributions jointly influence barrier-crossing kinetics may reveal additional scaling regimes and provide a more quantitative description of reaction dynamics in coarse-grained molecular systems. Extending the present analytical framework to such multi-component memory kernels therefore constitutes a promising avenue for future research. Our results demonstrate that oscillatory memory can qualitatively alter reaction kinetics beyond the well-established behavior of purely exponential memory kernels, highlighting the importance of accurately resolving oscillatory components when coarse-graining many-body dynamics.

\begin{acknowledgments}
We acknowledge support by Deutsche Forschungsgemeinschaft Grant CRC 1449 ”Dynamic Hydrogels at Biointerfaces”, Project ID 431232613, Project A03.
\end{acknowledgments}

\appendix
\section{Small auxiliary mass limit of $\Gamma_{\text{osc}}(t)$}
\label{sec.lim_osc-exp}
Here, we demonstrate how the oscillating memory kernel in eq.\eqref{eq.kern_osc} converges for small $m_y$ to a purely exponential kernel. Using eqs.(\ref{eq.taus_o_phi}), $\tau_\varphi$ can be rewritten as

\begin{equation}
\begin{split}
    \tau_\varphi &= \frac{1}{i\frac{\gamma}{2m_y} \sqrt{1-\frac{4k_y m_y}{\gamma^2}} } \\
    &=\frac{1}{i\frac{\gamma}{2m_y} \left(1-\frac{2k_y m_y}{\gamma^2} \right) } + \mathcal{O}(m_y^3).
    \end{split}
\label{eq.tau_phi_intermd}
\end{equation}

\noindent Inserting eq.(\ref{eq.tau_phi_intermd}) into eq.(\ref{eq.kern_osc}), we obtain

\begin{equation}
\begin{split}
       \Gamma_{\text{osc}}(t) 
    &\approx k_y  e^{- t / \tau_{\text{o}}} \left[ \cos\left(t\mathbin{/}\tau_\varphi \right) - i \sin\left(t\mathbin{/}\tau_\varphi \right) \right]\\
    &= k_y  e^{- t / \tau_{\text{o}} - it/\tau_\varphi},\\
    &\approx k_y  e^{- t k_y/\gamma},
    \label{eq.kernel_limit_osc_exp_interm}
\end{split}
\end{equation}

\noindent which is equivalent to eq.(\ref{eq.kernel_limit_osc_exp}). The oscillating kernel converges to an exponentially decaying kernel in the asymptotic limit of small auxiliary mass $m_y$ with decay time $\tau_{\text{a}}=\gamma / k_y$.

\medskip 
\bibliographystyle{apsrev4-2}

\bibliography{refs}

\addcontentsline{toc}{section}{References}

\end{document}


\newcommand{\beginsupplement}{
    \setcounter{section}{0}
    \renewcommand{\thesection}{S\arabic{section}}
    \setcounter{equation}{0}
    \renewcommand{\theequation}{S\arabic{equation}}
    \setcounter{table}{0}
    \renewcommand{\thetable}{S\arabic{table}}
    \setcounter{figure}{0}
    \renewcommand{\thefigure}{S\arabic{figure}}
    \newcounter{SIfig}
    \renewcommand{\theSIfig}{S\arabic{SIfig}}}

\title{Reaction dynamics in non-Markovian systems with non-monotonically decaying memory}
\author{~Artur Bakaev}
\author{~Otto Geburtig}
\author{~Benjamin A. Dalton}
\author{~Roland R. Netz}
\email{corresponding author. rnetz@physik.fu-berlin.de}
\affiliation{Freie Universit\"at Berlin, Fachbereich Physik, 14195 Berlin, Germany} 
\date{15. September 2026}

\maketitle

\tableofcontents
\setcounter{tocdepth}{12}
\setcounter{secnumdepth}{12}

\beginsupplement

\renewcommand{\theequation}{S\arabic{equation}} 
\renewcommand{\thefigure}{S\arabic{figure}}
\renewcommand\thesection{\arabic{section}}

\clearpage
\newpage

\newcommand{\twome}[1]{\textit{\textcolor{purple}{#1}}}
\newcommand{\form}[1]{\textcolor{blue}{#1}}
\newcommand{\twoB}[1]{\textcolor{red}{#1}}

\section{GLE simulation methods}

\subsection{Markovian embedding for oscillatory-exponential memory kernel}
Here we demonstrate the equivalence of the GLE in eq.(2) in the main text and the Markovian embedding in eq.(4) in the main text \cite{ayaz2022self,flo_pai}.

We start by rewriting eqs.(4c,4d) of the main text for the auxiliary particle in vectorial form and then solve that inhomogeneous differential matrix equation. For brevity, we introduce two timescales 

\begin{align}
\tau_1=\frac{m_y}{\gamma} && \tau_2=\frac{\gamma}{k_y},
\label{eq.embed_tau_a_b}
\end{align}

\noindent so that eqs.(4c,4d) of the main text become

\begin{align}
    \dot{y}(t) &= w(t)\\
    \dot{w}(t) &= \frac{-1}{\tau_1\tau_2}[x(t)-y(t)] - \frac{1}{\tau_1}w(t) + \frac{F_y(t)}{m_y}    .
\end{align}

\noindent Rewriting these equations in matrix form gives

\begin{eqnarray}
    \dot{\vec{z}}(t) = \textbf{A} \vec{z}(t) + \vec{\psi}(t),
    \label{eq.embed_vec1}
\end{eqnarray}

\noindent with 

\begin{align}
    \vec{z}(t) &= \mathlarger{\begin{pmatrix} y(t)  \\ w(t) \end{pmatrix}}, \label{eq.embed_vec_y}\\[2mm]
    \textbf{A} &= \mathlarger{\mathlarger{\begin{bmatrix} 0 & 1 \\ \frac{-1}{\tau_1\tau_2} & \frac{-1}{\tau_1} \end{bmatrix} }}, \\[2mm]
    \textbf{A}^{-1} &= \mathlarger{\mathlarger{\begin{bmatrix} \frac{-1}{\tau_1} & \frac{+1}{\tau_1\tau_2} \\ -1 & 0 \end{bmatrix} }}, \\[2mm]
    \vec{\psi}(t) &= \mathlarger{\begin{pmatrix} 0  \\ \frac{-x(t)}{\tau_1\tau_2} + \frac{F_y(t)}{m_y}  \end{pmatrix}}.
\end{align}

\noindent Eq.(\ref{eq.embed_vec1}) is an inhomogeneous differential matrix equation, whose solution is given by

\begin{equation}
    \begin{split}
        \vec{z}(t) &=  \exp\{ \textbf{A}(t-t_0)\} \vec{z}(t_0)  + \int^t_{t_0}  \exp\{ \textbf{A}(t-t') \} \vec{\psi}(t') dt'.
        \end{split}
   \label{eq.embed_sol}
\end{equation}

\noindent Next, we perform a partial integration for the term involving $x(t')$ in $\vec{\psi}(t')$

\begin{equation}
    \begin{split}
           \vec{z}(t) &=   \exp\{ \textbf{A}(t-t_0) \}   \vec{z}(t_0)        - \biggl \lceil  \textbf{A}^{-1} \exp\{ \textbf{A}(t-t') \} \begin{pmatrix} 0  \\ \frac{-x(t')}{\tau_1\tau_2}  \end{pmatrix}  \biggr \rceil_{t_0}^t \\
           &+  \int^t_{t_0} \textbf{A}^{-1} \exp\{ \textbf{A}(t-t') \} \begin{pmatrix} 0  \\ \frac{-v(t')}{\tau_1\tau_2}  \end{pmatrix} dt'  +\int^t_{t_0} \exp\{ \textbf{A}(t-t') \} \begin{pmatrix} 0  \\ \frac{F_y(t')}{m_y}  \end{pmatrix} dt'.
           \label{eq.embed_PI_sol}
    \end{split}
\end{equation}

\noindent To evaluate the matrix exponential, we diagonalize the matrix $\textbf{A}$ and write the eigenvalues $\lambda_{1,2}$ of  $\textbf{A}$ as 

\begin{align}
    \lambda_{1,2} = - \frac{1}{2 \tau_1} \pm \varphi , && \varphi = \sqrt{\frac{1}{4 \tau_1^2} - \frac{1}{\tau_1\tau_2}}.
    \label{eq.embed_eigenval}
\end{align}

\noindent The eigenvectors are given by

\begin{align}
    \vec{v}_1 =    \begin{pmatrix} -1  \\ -\lambda_1 \end{pmatrix}   ,        &&     \vec{v}_2 =    \begin{pmatrix} \lambda_2 + \frac{1}{\tau_1}  \\ -\frac{1}{\tau_1\tau_2}  \end{pmatrix}.
    \label{eq.embed_eigenvec}
\end{align}

\noindent The matrix exponential in eq.(\ref{eq.embed_PI_sol}) follows as 

\begin{equation}
    \begin{split}
        \exp\{ \textbf{A}t \} &= \frac{1}{\frac{1}{\tau_1 \tau_2}
+ \lambda_1\left(\lambda_2 + \frac{1}{\tau_1}\right)}\begin{pmatrix}
-1 & \lambda_2 + \frac{1}{\tau_1} \\
-\lambda_1 & -\frac{1}{\tau_1 \tau_2}
\end{pmatrix}
\begin{pmatrix}
e^{\lambda_1} & 0 \\
0 & e^{\lambda_2}
\end{pmatrix}
\begin{pmatrix}
-\frac{1}{\tau_1 \tau_2} & -\left(\lambda_2 + \frac{1}{\tau_1}\right) \\
\lambda_1 & -1
\end{pmatrix} \\
        &= \frac{1}{\lambda_2 - \lambda_1}   
        \mathlarger{\begin{bmatrix} \lambda_2 e^{\lambda_1 t} - \lambda_1 e^{\lambda_2 t} & e^{\lambda_2 t} - e^{\lambda_1 t}  \\ 
       \lambda_2 \lambda_1 (e^{\lambda_1 t} -  e^{\lambda_2 t} ) & \lambda_2 e^{\lambda_2 t} - \lambda_1 e^{\lambda_1 t}  \end{bmatrix}}\\[3mm]
        &=\exp\Bigl\{ \frac{-t }{ 2\tau_1} \Bigr\}
        \mathlarger{\begin{bmatrix} \cosh(\varphi t) + \frac{\sinh(\varphi t)}{2\tau_1 \varphi} & -\frac{\sinh(\varphi t)}{\varphi}  \\ 
       \frac{\sinh(\varphi t)}{\tau_1 \tau_2 \varphi} & \cosh(\varphi t) - \frac{\sinh(\varphi t)}{2 \tau_1  \varphi}  \end{bmatrix}},
    \end{split}
    \label{eq.embed_matrix1}
\end{equation}

\noindent and from that

\begin{equation}
\begin{split}
\textbf{A}^{-1}\exp\{ \textbf{A}t \} &= 
\tau_1\tau_2 \exp\Bigl\{ \frac{-t }{ 2\tau_1} \Bigr\}        \\
&\mathlarger{
\begin{bmatrix}
- \frac{\cosh(\varphi t)}{\tau_1} + \frac{\sinh(\varphi t)}{2\tau^2_a \varphi} +\frac{\sinh(\varphi t)}{\tau_1\tau_2\varphi} 
& -\frac{\sinh(\varphi t)}{2 \tau_1\varphi} -  \cosh(\varphi t) \\ 
\frac{\cosh(\varphi t)}{\tau_1\tau_2} + \frac{\sinh(\varphi t)}{\tau_1\tau_2 \varphi}
&  \frac{\sinh(\varphi t)}{ \tau_1 \tau_2 \varphi} 
\end{bmatrix}}.
\end{split}
\label{eq.embed_matrix2}
\end{equation}

\noindent Using eqs.(\ref{eq.embed_matrix1},\ref{eq.embed_matrix2}), the solution for $y(t)$ follows as the first entry of the vector $\vec{z}(t)$, as $y(t) =\vec{z}_1(t) $. Additionally, assuming $t_0 \rightarrow -\infty$ we obtain

\begin{equation}
\begin{split}
y(t) &= x(t) - \int^t_{-\infty} v(t') e^{-\frac{t-t'}{2\tau_1}} \left[ \cosh(\varphi[t-t']) + \frac{\sinh(\varphi[t-t'])}{2\tau_1\varphi}\right] \\[2mm]
& -  F_y(t') e^{-\frac{t-t'}{2\tau_1}} \frac{\sinh(\varphi[t-t'])}{\tau_1\varphi} \; dt' .
\end{split}
\label{eq.embed_sol_y}
\end{equation}

\noindent By inserting the solution for $y(t)$, eq.(\ref{eq.embed_sol_y}), into eq.(4b) of the main text, we obtain

\begin{equation}
    m\dot{v}(t) = - \nabla U[x(t)] - \int^\infty_0  \Gamma(t) v(t-t') dt'  + F_R(t),
\end{equation}

\noindent with the memory kernel and random force given by

\begin{align}
    \Gamma(t) &= k e^{-\frac{t}{2\tau_1}} \left[ \cosh(\varphi t) + \frac{\sinh(\varphi t)}{2\tau_1\varphi}\right] , \label{eq.embed_kern_hyperb} \\[3mm]
    F_R(t) &= \int^\infty_0  F_y(t-t')e^{\frac{-t'}{2\tau_1}} \frac{\sinh(\varphi t')}{\tau_1\varphi} \; dt' .\label{eq.embed_F_R_hyperb}
\end{align}

\noindent Rewriting $\varphi$ as
\begin{equation}   
    \varphi = i \sqrt{\frac{k_y}{m_y} - \frac{\gamma^2}{4m_y^2 } }
    \label{eq.embed_intro_tauPhi}
\end{equation}

\noindent the kernel in eq.(\ref{eq.embed_kern_hyperb}) can, using $\tau_1$ and $\tau_2$, eqs.(\ref{eq.embed_tau_a_b}), be expressed as

\begin{equation}
  \Gamma(t) = k_y   e^{- t \frac{\gamma}{2m_y}} \left[ \cos\left(t\sqrt{\frac{k_y}{m_y} - \frac{\gamma^2}{4m_y^2 } } \right) + \frac{\gamma}{2m_y \sqrt{\frac{k_y}{m_y} - \frac{\gamma^2}{4m_y^2 } } } \sin\left(t\sqrt{\frac{k_y}{m_y} - \frac{\gamma^2}{4m_y^2 } } \right) \right] .
  \label{eq.kern_osc_kmg}
\end{equation}

\noindent Finally, by introducing the exponential decay time $\tau_{\text{o}}$ and the oscillation time $\tau_\varphi$ as

\begin{align}
    \tau_{\text{o}} = \frac{2m_y}{\gamma} , && \tau_\varphi = \frac{1}{\sqrt{\frac{k_y}{m_y} - \frac{\gamma^2}{4m_y^2}}},
    \label{eq.taus_O_phi}
\end{align}

\noindent we obtain the kernel in the form of eq.(3) of the main text.

\subsection{Dimensionless Markovian embedding}

By rescaling position $x$ and time $t$ by the length scale $L$ and the diffusion time $\tau_{\text{D}}$ as $x=\tilde{x}L$ and    $t=\tilde{t}\tau_{\text{D}}$, the Markovian embedding, eqs.(4) in the main text, becomes dimensionless

\begin{subequations}
\begin{align}
   \dot{\tilde{x}}\left(\tilde{t}\right) &= \tilde{v}\left(\tilde{t}\right) \\ 
    \frac{\tau_{\text{m}}}{\tau_{\text{D}}}\dot{\tilde{v}}\left(\tilde{t}\right) &= \frac{\tau_{\text{D}}}{\gamma} k_y[\tilde{x}(\tilde{t})- \tilde{y}(\tilde{t})] 
    - \beta L \nabla U[L\tilde{x}(\tilde{t})]\label{eq:ham_eqs_extend_v_dim}\\
    \dot{\tilde{y}}(\tilde{t}) &= \tilde{w}(\tilde{t}) \label{eq:ham_eqs_extend_y}\\
    \frac{\tau_{m_y}}{\tau_{\text{D}}} \dot{w}(t) &= -\frac{\tau_{\text{D}}}{\gamma} k_y[\tilde{x}(\tilde{t})- \tilde{y}(\tilde{t})] - \tau_{\text{D}} \tilde{w}(\tilde{t}) + \tilde{F}_y(\tilde{t}).\label{eq:ham_eqs_extend_w_dim}
\end{align}
\label{eq:ham_eqs_extend_dimless}
\end{subequations}

\noindent Here, $\tilde{y}=y/L$ is the rescaled auxiliary variable and  $\tilde{F}_y = \beta L F_y$ is the rescaled random force with zero mean and variance $\langle  \tilde{F}_y(\tilde{t})\tilde{F}_y(\tilde{t}') \rangle = 2 \delta(\tilde{t}-\tilde{t}')$.

\subsection{Numerical convergence analysis of GLE simulations with an oscillatory-exponential memory kernel}

In this section, the accuracy of the numerical simulations of the GLE with an oscillating kernel is discussed. The aim is to determine optimal choices of the simulation time step $\Delta$ and the simulation length $N_{\text{steps}}$, both with respect to the timescales of the system $\tau_{\text{m}}$, $\tau_\varphi , \tau_{\text{o}}$ relative to the system's diffusion time $\tau_{\text{D}}$. 
We compare three combinations of timescales $\tau_{\text{m}}, \tau_\varphi , \tau_{\text{o}}$, chosen so that for each combination, one short, one intermediate, and one long timescale is represented. 
To quantify the numerical convergence of the studied trajectories, we analyse two quantities: $R$, the root-mean-square deviation of the extracted potential, and $\tau_{\text{MFP}}$. 

$R$ is obtained by comparing the extracted potential $U^{\text{\text{extr}}}(x) \sim -\log[p(x)] k_BT$ with the input potential of the simulation $U(x)= U_0[(\frac{x}{L})^2 - 1]^2$, according to

\begin{equation}
R = \sqrt{\frac{1}{N_{\text{bins}}} \sum^{N_{\text{bins}}}_i \left[U^{\text{extr}}(x_i) - U(x_i) \right]^2 }.
\label{eq.RMS}
\end{equation}

$U^{\text{extr}}(x)$ is shifted vertically so that min$[U^{\text{extr}}(x)] = 0$. The distribution $p(x(t))$ is a normalized histogram of the full particle's trajectory $x(t)$ with $N_{\text{bins}}=100$ equidistant bins centred at each data point $x_i$.

\begin{figure}
\centering
    {{\includegraphics[width=0.9 \textwidth]{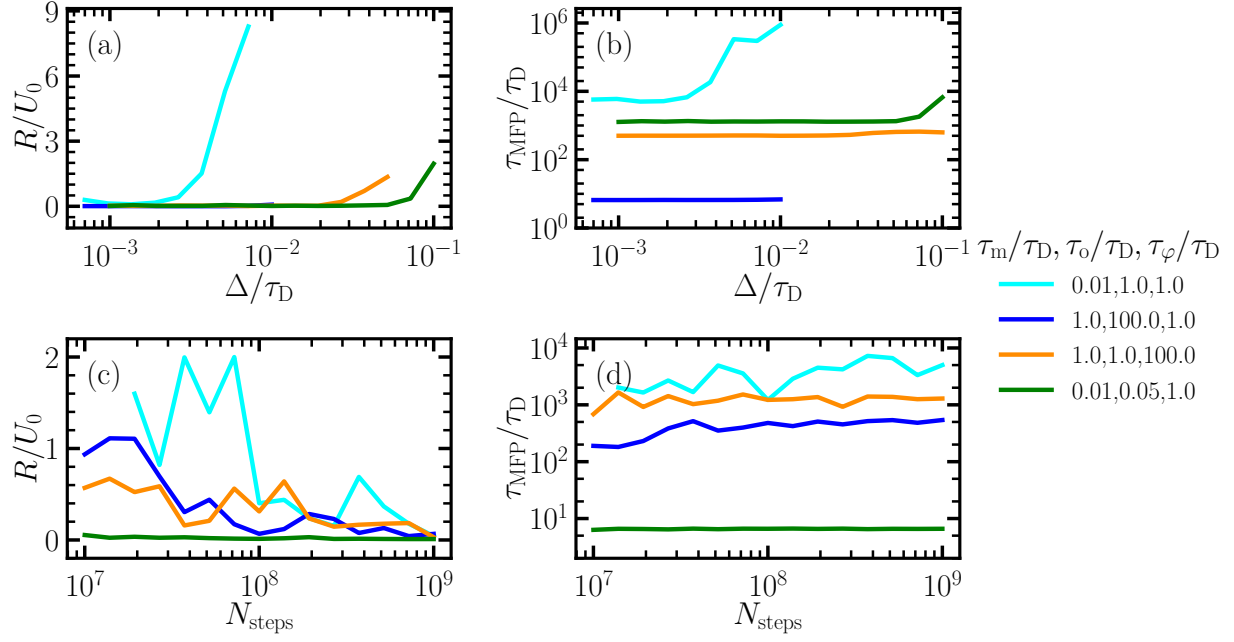} }}
    \caption{Numerical convergence of the root-mean-square error $R$, defined in eq.\eqref{eq.RMS}, and the mean first passage time $\tau_{\text{MFP}}$ as functions of time step ratio $\Delta/\tau_\text{D}$ and number of steps  $N_{\text{steps}}$ of the simulation. (a-b) Varying time step $\Delta$ while keeping the trajectory length constant at $T/\tau_\text{D}=5\cdot10^6$. (c,d) Varying simulations steps $N_\text{steps}$ while keeping the time step constant at $\Delta/\tau_\text{D}=10^{-3}$. } 
    \label{fig.conv}
\end{figure}

\subsubsection{Analysis of time step $\Delta$}
We run a set of simulations with varying time step ratios $\Delta/\tau_\text{D}$ while keeping the trajectory length $T/\tau_\text{D}= N_{\text{steps}} \Delta/\tau_\text{D}$ constant and calculate the quantities $R$ and $\tau_{\text{MFP}}$. We aim to determine the maximal $\Delta$ so that the error $R$ remains small and $\tau_{\text{MFP}}$ converges, see Figure \ref{fig.conv} (a-b). As expected, both $R$ and $\tau_{\text{MFP}}$ change with increasing time step ratio $\Delta/\tau_\text{D}$. We see that there is no universal choice of $\Delta$: whereas in some instances (cyan curves in Figure \ref{fig.conv} a-b), both the error $R$ and $\tau_{\text{MFP}}$ increase by more than 100\% for $\Delta/\tau_\text{D}$ approaching the smallest timescale ratio, in other instances (green curves in Figure \ref{fig.conv} a-b) the system remains robust even when $\Delta/\tau_\text{D}$ is larger than the shortest timescale. Thus, in our numerical simulations, we choose $\Delta/\tau_\text{D}$ to be one to two orders of magnitude smaller than the smallest timescale ratio of the system.

\subsubsection{Analysis of step number $N_{\text{steps}}$}

Here we run a set of simulations with varying $N_{\text{steps}}$ while keeping $\Delta/\tau_\text{D}=10^{-3}$. We aim to determine the trajectory length required for $R$ to remain small and for $\tau_{\text{MFP}}$ to be estimated reliably, see Figure \ref{fig.conv} c-d. 
As for the choice of $\Delta$, there is no universal value of $N_{\text{steps}}$ that guarantees convergence for all parameter sets. The deviation $R$ approaches small values at different rates for the different timescale ratios and remains small at $N_{\text{steps}}=10^9$. For three of the four parameter sets $\tau_{\text{MFP}}$ becomes approximately independent of $N_{\text{steps}}$ over the largest trajectory lengths considered (green, blue, and orange curves in Figure \ref{fig.conv} d). For the remaining parameter set (cyan curve), substantial statistical fluctuations persist up to $N_{\text{steps}}=10^9$, although no systematic dependence on $N_{\text{steps}}$ is apparent. Thus, while $N_{\text{steps}}\sim10^9$ is sufficient for the majority of the parameter sets considered here, longer trajectories may be required to obtain statistically stable estimates of  $\tau_{\text{MFP}}$ in regimes with particularly slow barrier-crossing dynamics.

\section{Different measures of barrier-crossing times for a harmonic potential}

\begin{figure}[h]
\centering
       \includegraphics[width=0.6\textwidth]{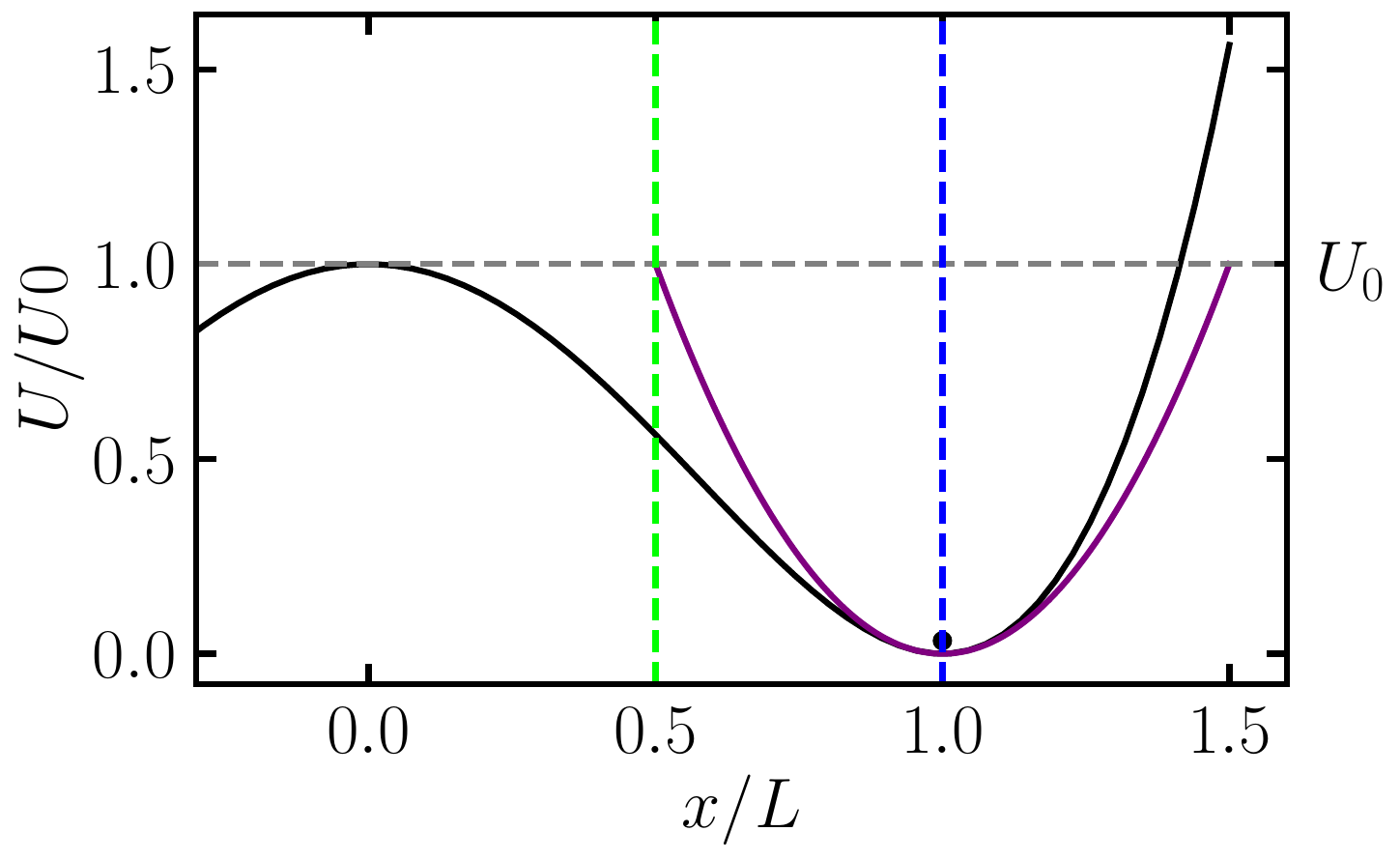}
        \caption{Double-well potential (black solid line) with barrier height $U_0$ and well separation $2L$, together with a harmonic approximation (purple line) with curvature $K = 8U_0/L^2$. Dashed lines indicate the barrier height (grey), the well minimum (blue), and the position where the harmonic approximation reaches $U_0$ (green).
        } 
    \label{fig.DW_HW}
\end{figure}

In this section, we compare different measures for quantifying the barrier-crossing times for GLE systems with oscillating memory. Whereas the mean all-to-first passage time $\tau_{\text{MFP}}$ (which is the quantity used in the main text) is defined as the average over all first-passage events, the first-to-first passage time, $\tau_{\text{MFFP}}$, is defined using only the time intervals between the first transition of a particle entering a state and the first transition into the other state (the time interval between the green and red crosses in Figure 1 b of the main text). Both $\tau_{\text{MFP}}$ and $\tau_{\text{MFFP}}$ are extracted from long trajectories. The escape time $\tau_{\text{esc}}$ is defined in terms of the survival probability $S(t)$ for a particle density $\rho(x,t)$

\begin{equation}
S(t) = \int\limits_{-\infty}^{x_\text{abs}} \rho(x,t) \mathrm{d}x.
\label{eq.Otto2.45}
\end{equation}

\noindent crossing from $-\infty$ to an absorbing boundary $x_\text{abs}$. The first-passage distribution $\rho_\tau(t)$ for the particles is the rate of change of $S(t)$

\begin{equation}
\rho_\tau(t) = -\frac{\mathrm{d}S(t)}{\mathrm{d}t}.
\label{eq.Otto2.46}
\end{equation}

\noindent The escape time $\tau_{\text{esc}}$ then is the first moment of $\rho_\tau(t)$:

\begin{equation}
\begin{split}
    \tau_{\text{esc}} 
    &= \int\limits_{0}^{\infty}  S(x_0,t) \mathrm{d}t .
    \end{split}
    \label{eq.otto2.47}
\end{equation}

\begin{figure}
   \centering 
    \includegraphics[width=0.5\textwidth]{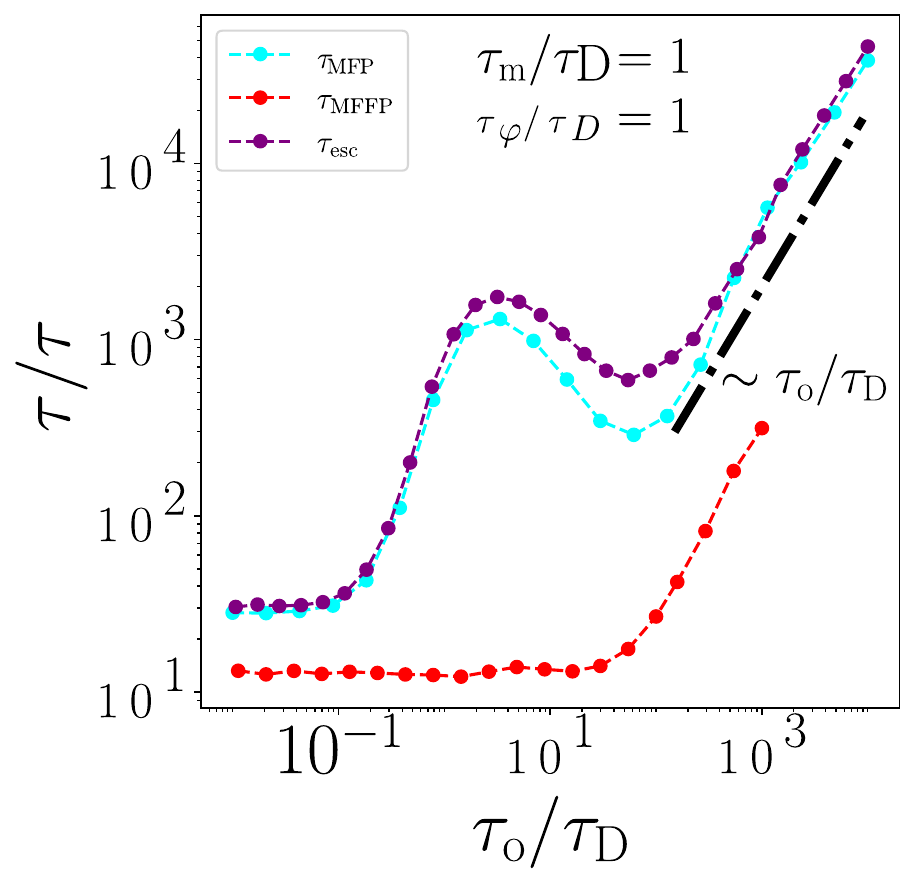}
    \caption{Comparison of the different measures of barrier crossing times $\tau_{\text{MFP}},\tau_{\text{MFFP}}$ and $\tau_{\text{esc}}$, all obtained for a particle in a harmonic well potential, eq.\eqref{eq.hw}, with oscillating memory, extracted from a trajectory obtained by integrating eqs.(4) of main text. The black dash-dotted line indicates scaling behaviour linear in $\tau_{\text{o}}/\tau_{\text{D}}$. Here, we fix $\tau_{\text{m}}/\tau_{\text{D}}=1$ and $\tau_\varphi/\tau_{\text{D}}=1$.}
    \label{fig.MAFPT_MFFPT_ESC}
\end{figure}

\noindent Here, we consider a single harmonic well
\begin{align}
U_{\text{hw}}(x) = \frac{K}{2}x^2, \qquad K = 8U_0/L^2,
\label{eq.hw}
\end{align}
with stiffness $K$ obtained by approximating the double-well potential, eq.(7) of the main text, at one well minimum. To obtain the escape time $\tau_{\text{esc}}$, the simulations are terminated once the particle reaches  $x_f/L=1/2$, such that $U_{\text{hw}}(x_f)=U_0$, see Figure \ref{fig.DW_HW} for an illustration.
The measure used in the main text, $\tau_{\text{MFP}}$, closely agrees with the escape time $\tau_{\text{esc}}$, see Figure \ref{fig.MAFPT_MFFPT_ESC}, as has been shown for systems with exponential memory before \cite{kappler2018, Recross_Ben}. In contrast,   $\tau_{\text{MFFP}}$ deviates significantly from both $\tau_{\text{MFP}}$ and $\tau_{\text{esc}}$:  $\tau_{\text{MFFP}}$ remains constant for intermediate decay times $\tau_{\text{o}}$ and scales linearly with $\tau_{\text{o}}$ for long $\tau_{\text{o}}$, missing the memory-induced slowdown captured by $\tau_{\text{MFP}}$ and $\tau_{\text{esc}}$ for intermediate decay times $\tau_{\text{o}}$. Unlike systems with purely exponential memory, $\tau_{\text{MFFP}}$ for oscillating memory captures the memory-induced slowdown for long decay times $\tau_{\text{o}}$ \cite{Recross_Ben}.

\section{Deriving $\tau_{\text{MFP}}$ for different potentials}

Performing a saddle-point approximation for the double integral in eq.(11) of the main text yields

\begin{equation}
    \tau_{\text{MFP}} = \tau_\text{rel} \Theta e^{\beta U_0},
    \label{eq.mfpt_rel_theta}
\end{equation}

\noindent with the prefactor $\Theta$

\begin{subequations}
\begin{empheq}[left={ \Theta = \empheqlbrace }]{align}
    & \sqrt{\frac{\pi}{\beta U_0}}, & \text{harmonic well potential} \label{eq.theta_hw} \\
    & 2\pi \sqrt{\frac{K_\text{min}}{K_\text{bar}}}, & \text{double well potential} \label{eq.theta_dw}
\end{empheq}
\end{subequations}

\noindent depending on the potential under consideration \cite{Netz26}. The simplest potential giving rise to barrier-crossing dynamics is a symmetric double-well potential $U(x)$ (see eq.(7) of main text) characterized by barrier height 
$U_0$, well separation $2L$ and curvature magnitudes at the minima and at the barrier, $K_{\text{min}}$ and $K_{\text{bar}}$ respectively, given by

\begin{align}
    K_{\text{min}} &= 8U_0 / L^2, &&     K_{\text{bar}} = 4U_0/L^2.
    \label{eq.Uprimes}
\end{align}
For the two-point correlation function $C(t) = \langle x(t) x(0)\rangle $, we obtain in the harmonic approximation 
\begin{equation}
C_0=\frac{k_BT}{K_\text{min}}.
\label{eq.C_0}
\end{equation}

\noindent When using eq.(13) in the main text for calculating $\tau_\text{rel}$ with a correlation function extracted from simulated trajectories in a double well potential, $C_\text{sim}(t)$, we use the extracted value for $C_\text{sim}(t=0)$, and obtain the grey dash-dotted lines in Figure 5 b-c in the main text.

\section{Deriving $\tau_{\text{MFP}}$ for GLE systems with oscillatory-exponential memory kernel}
\label{sec.deriving_tau_rel}
In this section, we present the explicit solution for the relaxation time, eq.(13) in the main text. The GLE eq.(2) in the main text can for a harmonic potential be Fourier transformed according to $\tilde{x}(\omega) = \int dt \, e^{-i\omega t} x(t)$, yielding
\begin{eqnarray}
    \tilde{x}(\omega) = \tilde{\chi}(\omega) \tilde{F}_R(\omega).
\end{eqnarray}

\noindent Here, $\tilde{F}_R(\omega)$ denotes the random force, and the response function $\tilde{\chi}(\omega)$ is given by
\begin{equation}
    \tilde{\chi}(\omega) = [K - m\omega^2 + i \omega \tilde{\Gamma}^+(\omega)]^{-1},
\end{equation}
\noindent for a harmonic potential with stiffness $K$ and half-sided Fourier transform of the memory kernel $\tilde{\Gamma}^+(\omega)=\int_0^{\infty} dt \, e^{-i\omega t} \Gamma(t)$. The positional correlation function $\tilde{C}(\omega)$ is then given by
\begin{eqnarray}
    \begin{split}
       \beta \tilde{C}(\omega) &= \left(\tilde{\Gamma}^+(\omega) + \tilde{\Gamma}^+(-\omega) \right) \tilde{\chi}(\omega) \tilde{\chi}(-\omega)\\
       &=- \frac{1}{i \omega} \left(\tilde{\chi}(\omega) - \tilde{\chi}(-\omega)\right).
    \end{split}
\end{eqnarray}
\noindent Since the position--velocity correlation function $C_{xv}(t)$ is related to the positional correlation function $C(t)$ via
\begin{equation}
    \begin{split}
C_{xv}(t) &= \frac{d}{dt}C(t) \\
&= \int \frac{d\omega}{2\pi} e^{i\omega t} \, i\omega \tilde{C}(\omega),
    \end{split}
\end{equation}
\noindent it follows that
\begin{align}
\beta \tilde{C}_{xv}(\omega) = \tilde{\chi}(-\omega) - \tilde{\chi}(\omega).
\end{align}
\noindent Transforming back to the time domain yields
\begin{align}
C(t) = - \int_{t}^{\infty} ds \, C_{xv}(s).
\label{eq.C_xx_Cvx}
\end{align}

\noindent Since $\tilde{C}_{xv}(\omega)$ is odd, the full correlation function is determined by $t>0$. The single-sided position--velocity correlation function gives $\tilde{C}^+_{xv}(\omega) = -\tilde{\chi}(\omega)$, and substituting into eq.\eqref{eq.C_xx_Cvx}, the single-sided positional correlation function $C^+(t)$ can then be written as
\begin{align}
\beta C^+(t) = \Theta(t) \int_{t}^{\infty} \chi(s)\, ds .
\label{eq.corr_chi}
\end{align}

\noindent The memory kernel $\Gamma_{\mathrm{osc}}(t)$, with $k_y= \frac{\gamma \tau_{\text{o}}}{2} \left(\frac{1}{\tau_{\text{o}}^2}+ \frac{1}{\tau_\varphi^2} \right)$, has the Fourier transform 

\begin{equation} 
\begin{split}
\tilde{\Gamma}^+_\text{osc}(\omega) & = \int_{0}^{\infty} dt e^{-i \omega t} \Gamma_\text{osc}(t) \\
& = k_y \frac{ 2 / \tau_{\text{o}}  + i\omega } { 1/\tau_\varphi^2 + \left(1/\tau_{\text{o}} + i \omega \right)^2 }.
 \label{eq.half_kern_osc_FT}
\end{split}
\end{equation}

\noindent Introducing $\aleph^2=\frac{1}{\tau_{\text{o}}^2}+ \frac{1}{\tau_\varphi^2}$ for shorthand notation, the response function $\tilde{\chi}(\omega)$ then can be expressed as 

\begin{equation}
\begin{split}
    \tilde{\chi}(\omega) &= [K - m\omega^2 + i \omega \tilde{\Gamma}^+_\text{osc}(\omega)]^{-1}    \\
    &= \frac{1/\tau_\varphi^2 + \left(1/\tau_{\text{o}} + i \omega \right)^2 }{m[\omega^4 - 2i/\tau_{\text{o}} \; \omega^3 - \left(\frac{K+k_y}{m} + \aleph^2 \right)\omega^2 + 2i/\tau_{\text{o}}\left(\frac{K+k_y}{m}\right)\omega + \frac{K}{m}\aleph^2]}.
\end{split}
\label{eq.chi_om}
\end{equation}

\noindent Eq.\eqref{eq.chi_om} can be Fourier transformed into the time domain as 

\begin{equation}
    \begin{split}
        \chi(t) &= \int_{-\infty}^{\infty} \frac{d\omega}{2\pi} e^{i\omega t} \tilde{\chi}(\omega) \\
        &= 2\pi i \sum_k \text{Res}[\frac{e^{i\omega t}}{2\pi} \tilde{\chi}(\omega),\omega_k]\\
        &= i \sum_{k=1}^4 \frac{1/\tau_\varphi^2 + \left(1/\tau_{\text{o}} + i \omega_k \right)^2}{m \prod\limits_{j\neq k} (\omega_k - \omega_j)} e^{i \omega_k t}.
    \end{split}
    \label{eq.chi_t}
\end{equation}

\noindent Here, we introduce the poles $\omega_k$ of $\tilde{\chi}(\omega)$.

\subsection{Poles of $\tilde{\chi}(\omega)$}
In eq.\eqref{eq.chi_t}, the 4 roots $\omega_k$ of the denominator of $\tilde{\chi}(\omega)$ appear. We use the Euler method \cite{Euler} for obtaining those, that constructs the four roots $\omega_k$ of the original quartic polynomial by obtaining the three roots $z_k$ of a corresponding resolvent cubic polynomial. We start with the quartic polynomial

\begin{align}
\omega^4 - 2i/\tau_{\text{o}} \; \omega^3 - \left(\frac{K+k_y}{m} + \aleph^2 \right)\omega^2 + 2i/\tau_{\text{o}}\left(\frac{K+k_y}{m}\right)\omega + \frac{K}{m}\aleph^2 &= 0 \label{eq.w_orig} \\ 
\omega^4 + a_3 \omega^3 +a_2 \omega^2 + a_1 \omega + a_0 &= 0,
\end{align}

\noindent and derive the reduced quartic polynomial by the substitution $\omega=y-a_3/4$,

\begin{equation}
y^4 + b_2 y^2 + b_1 y + b_0 = 0.
\label{eq.y_depr4}
\end{equation}
Next, we substitute

\begin{eqnarray}
    y^\mp= \pm \sqrt{z_1} \pm \sqrt{z_2} \pm \sqrt{z_3}
    \label{eq.y_subst}
\end{eqnarray}
in eq.\eqref{eq.y_depr4}, which yields



\begin{equation}
\begin{split}
4 \left(z_2 z_3 + z_3 z_1 + z_1 z_2 \right)
+ \left(z_1 + z_2 + z_3 \right)^2
+ b_2 \left(z_1 + z_2 + z_3 \right) + b_0\\
+ \left(\sqrt{z_2 z_3} + \sqrt{z_3 z_1} + \sqrt{z_1 z_2} \right)
\left[4 \left(z_1 + z_2 + z_3 \right) + 2b_2 \right]
+ \left(\sqrt{z_1} + \sqrt{z_2} + \sqrt{z_3} \right)
\left[8 \sqrt{z_1 z_2 z_3} \pm b_1 \right]
= 0  
\end{split}
\label{eq.y_subst}
\end{equation}

\noindent Eq.\eqref{eq.y_subst} is satisfied when

\begin{align}
    \sqrt{z_1z_2z_3 } =\mp \frac{b_1}{8} ,&& z_1 + z_2 + z_3 = -\frac{b_2}{2}, && z_2 z_3 + z_3 z_1 + z_1 z_2 = \frac{b_2^2 - 4b_0}{16}.
    \label{eq.z_cond_resolvent}
\end{align}


\noindent The relations in eq.\ref{eq.z_cond_resolvent} are the elementary polynomials defining the roots of the following cubic polynomial

\begin{equation}
z^3 + c_2 z^2 + c_1 z + c_0 = 0,
\label{eq.z_res}
\end{equation}

\noindent where the coefficients $c_j$ are related to the coefficients $b_j$ by eq.\eqref{eq.coeffs}. Eq.\ref{eq.z_res} is the resolvent cubic polynomial \cite{Euler} associated with eq.\eqref{eq.y_depr4}. The roots of eq.\eqref{eq.y_depr4} can therefore be constructed from the roots of the resolvent cubic eq.\eqref{eq.z_res}. The  form of the roots of eq.\eqref{eq.y_depr4} follows from the condition


\begin{equation} 
    \sqrt{z_1z_2z_3} = \mp \frac{b_1}{8}
    \label{eq.pm_condit}
\end{equation}

\noindent as

\begin{equation} 
\begin{aligned}
y^{-}_1 &= \sqrt{z_1} + \sqrt{z_2} + \sqrt{z_3}, & \quad & y^{+}_1 = -\sqrt{z_1} - \sqrt{z_2} - \sqrt{z_3}, \\
y^{-}_2 &= \sqrt{z_1} - \sqrt{z_2} - \sqrt{z_3}, & \quad & y^{+}_2 =  -\sqrt{z_1} + \sqrt{z_2} + \sqrt{z_3},\\
y^{-}_3 &= -\sqrt{z_1} + \sqrt{z_2} - \sqrt{z_3}, & \quad & y^{+}_3 = +\sqrt{z_1} - \sqrt{z_2} + \sqrt{z_3}, \\
y^{-}_4 &= -\sqrt{z_1} - \sqrt{z_2} + \sqrt{z_3}, & \quad & y^{+}_4 = +\sqrt{z_1} + \sqrt{z_2} - \sqrt{z_3}, \\
\end{aligned}
\label{eq.y_sol}
\end{equation}

\noindent since both sets of solutions in eq.\eqref{eq.y_sol} solve eq.\eqref{eq.y_subst}. 
For solving eq.\eqref{eq.z_res}, the resolvent polynomial is reduced by substituting $z = t - c_2/3$,

\begin{equation}
t^3 + d_1 t + d_0 = 0.
\label{eq.t_depr3}
\end{equation}

\noindent Eq.\eqref{eq.t_depr3} can be solved by yet another substitution $t = u - \frac{d_1}{3u}$

\begin{equation}
u^3  - \frac{d_1^3}{27u^3} +d_0 = 0,
\label{eq.u_subs}
\end{equation}

\noindent which yields a polynomial quadratic in $u^3 = x$

\begin{equation}
x^2 + d_0 x - d_1^3/27 = 0
\label{eq.x_quad}
\end{equation}

\noindent with solutions in $x$

\begin{equation}
x_\pm = - \frac{d_0}{2} \pm \sqrt[2]{ \frac{d_0^2}{4} + \frac{d_1^3}{27}}.
\label{eq.x_quad_sol}
\end{equation}

\noindent One can show that the two solutions $x_\pm$ yield equivalent solutions for $u_k$, therefore without loss of generality, one can write the three solutions in $u$ as

\begin{equation}
u_k = \epsilon_k \sqrt[3]{ - \frac{d_0}{2} + \sqrt[2]{ \frac{d_0^2}{4} + \frac{d_1^3}{27}}},
\label{eq.u_sol}
\end{equation} 

\noindent using the three cubic roots of unity $\sqrt[3]{+1}  = \epsilon_k=\{1,-\frac{1}{2} \pm i\frac{\sqrt[2]{3}}{2}\}$, for which the relations

\begin{equation}
    \begin{aligned}
        \epsilon_2\epsilon_3 &=1, & \epsilon_2^2 &=\epsilon_3, & \epsilon_3^2 &=\epsilon_2\\
        \epsilon_2 &= \frac{1}{\epsilon_3}, & \epsilon_3 &= \frac{1}{\epsilon_2}, & \epsilon_2 + \epsilon_3 &= -1 
    \end{aligned}
\end{equation}

\noindent hold. We thus obtain the roots $z_k$ in terms of $u_k$ as

\begin{equation}
z_k = t_k - c_2/3 = u_k - \frac{d_1}{3 u_k} - c_2/3.
\label{eq.z_sol}
\end{equation}

\noindent Having these three solutions $z_k$, we construct the four solutions of $y_k$ according to eq.\eqref{eq.y_sol}.
With $\omega_k^\mp = y_k^\mp - a_3/4$, the original roots are given by

\begin{equation}
\begin{split}
\omega_1^\mp &= \pm (\sqrt{z_1} + \sqrt{z_2} + \sqrt{z_3}) + \frac{i}{2\tau_{\text{o}}},\\
\omega_2^\mp &= \pm (\sqrt{z_1} - \sqrt{z_2} - \sqrt{z_3}) + \frac{i}{2\tau_{\text{o}}},\\
\omega_3^\mp &= \pm (-\sqrt{z_1} + \sqrt{z_2} - \sqrt{z_3}) + \frac{i}{2\tau_{\text{o}}},\\
\omega_4^\mp &= \pm (-\sqrt{z_1} - \sqrt{z_2} + \sqrt{z_3}) + \frac{i}{2\tau_{\text{o}}}.\\
\end{split}
\label{eq.w_i_expl}
\end{equation}

\noindent  The various coefficients are given by

\begin{equation}
\begin{aligned}
    a_3 &= -\frac{2i}{\tau_{\text{o}}}, & a_2 &= -\left(\frac{K+k_y}{m} + \aleph^2\right), \\
    a_1 &= \frac{2i}{\tau_{\text{o}}}\left(\frac{K+k_y}{m}\right), & a_0 &= \frac{K}{m}\aleph^2, \\
    b_2 &= -\frac{3 a_3^2}{8} + a_2, & b_1 &= \frac{a_3^3}{8} - \frac{a_3 a_2}{2} + a_1, \\
    b_0 &= \frac{-3 a_3^4}{256} + \frac{a_3^2 a_2}{16} - \frac{a_3 a_1}{4} + a_0, \\
    c_2 &= \frac{b_2}{2}, & c_1 &= \frac{b_2^2 - 4b_0}{16}, &   c_0 &= -\frac{b_1^2}{64}, \\
    d_1 &= c_1 - \frac{c_2^2}{3}, &    d_0 &= \frac{2c_2^3}{27} - \frac{c_1 c_2}{3} + c_0.
\end{aligned}
\label{eq.coeffs}
\end{equation}

\subsection{Derivation of the correlation function and the relaxation time $\tau_{\text{rel}} $}
\noindent In terms of the poles $\omega_k$ and combining eqs.(\ref{eq.corr_chi},\ref{eq.chi_t}), the correlation function  is given by

\begin{equation}
\begin{split}
\beta C(t) &=-  \sum_{k=1}^4 \frac{ 
1/\tau_\varphi^2 + \left(1/\tau_{\text{o}} + i \omega^\mp_k \right)^2 
}
{m \prod\limits_{j\neq k} (\omega_k^\mp - \omega_j^\mp)} \frac{e^{i \omega^\mp_k t}}{\omega^\mp_k} \\
&= \frac{\sum_k^4 (-1)^k  
f^\pm_k (\omega^\mp_r - \omega^\mp_s) (\omega_r^\mp - \omega^\mp_u) (\omega_s^\mp - \omega_u^\mp) \omega_r^\mp \omega_s^\mp \omega_u^\mp
}{m\prod\limits_{l < o} (\omega_l^\mp - \omega_o^\mp) \prod\limits_{k} \omega_k^\mp}  e^{i \omega_k^\mp t}, \text{with $r<s<u \in \{1,2,3,4\}$} ,
\label{eq.C_t}
\end{split}
\end{equation}

\noindent where we introduce for shorthand notation:

\begin{align}
 f^\mp_k &=   1/\tau_\varphi^2 + \left(1/\tau_{\text{o}} + i \omega^\mp_k \right)^2,\\
 g^\mp_k &= f^\pm_k (\omega^\mp_r - \omega^\mp_s) (\omega_r^\mp - \omega^\mp_u) (\omega_s^\mp - \omega_u^\mp) \omega_r^\mp \omega_s^\mp \omega_u^\mp.
\end{align}

\noindent Then, with

\begin{equation}
    \begin{split}
        &\prod\limits_{k} \omega^\mp_k = \frac{K}{m}\aleph^2 , \\
        &\prod\limits_{l < o} (\omega^\mp_l - \omega^\mp_o) = 64 (z_1 - z_2)(z_1 - z_3)(z_2 - z_3),
    \end{split}
    \label{eq.relations_prod_omega}
\end{equation}

\noindent the correlation function can be written as

\begin{equation}
 C(t) =k_BT \frac{ \sum_k^4 (-1)^k g^\mp_k  e^{i \omega_k^\mp t} }{64 (z_1 - z_2)(z_1 - z_3)(z_2 - z_3) K \aleph^2}.
\end{equation}

\noindent In eq.\eqref{eq.relations_prod_omega}, note that the phase $\mp1$ from eq.\eqref{eq.pm_condit} appears in even orders, $(\mp1)^4=1$ and $(\mp1)^6=1$. We verify that this expression is consistent with the equipartition theorem, i.e., $C(t=0)=C_0=k_BT/K$

\begin{equation}
    \begin{split}
        \beta C(t=0) &=\frac{-g^\mp_1+g^\mp_2-g^\mp_3+g^\mp_4}{64 (z_1 - z_2)(z_1 - z_3)(z_2 - z_3) K \aleph^2} \\
        &= \frac{64 (z_1 - z_2)(z_1 - z_3)(z_2 - z_3) \aleph^2 }{64  (z_1 - z_2)(z_1 - z_3)(z_2 - z_3) K\aleph^2} \\
        &= \frac{1}{K}.
    \end{split}
\end{equation}

\begin{figure}
   \centering 
    \includegraphics[width=1\textwidth]{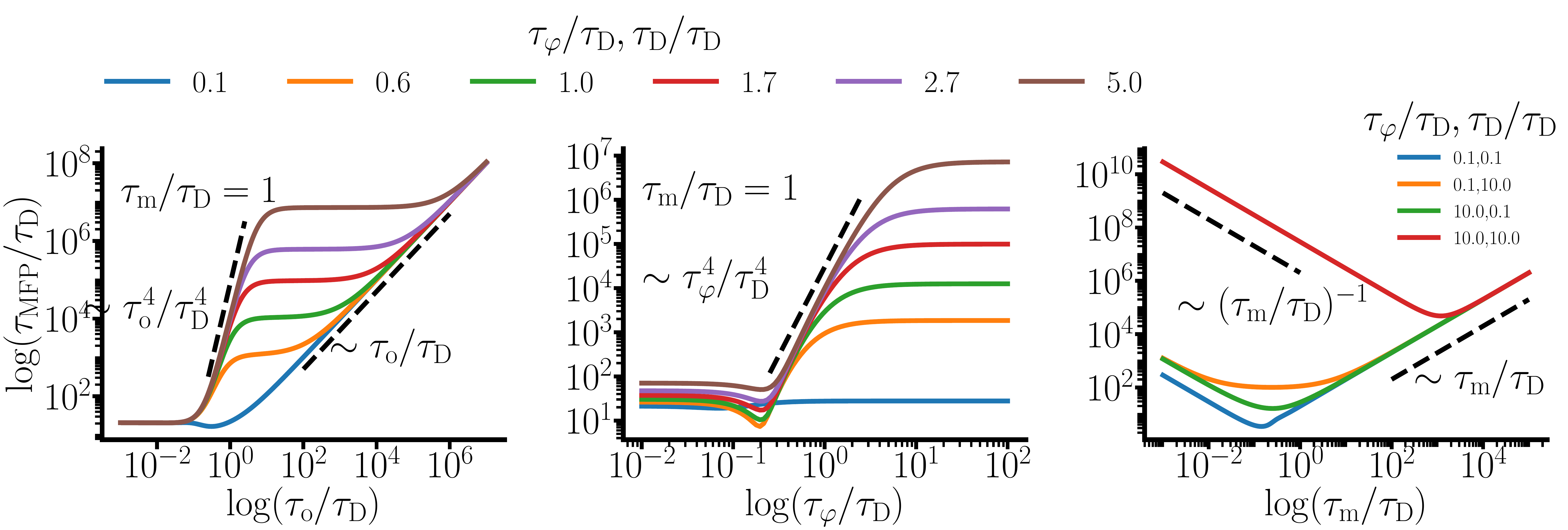}
    \caption{Analytical prediction of $\tau_{\text{MFP}}$ in a double well potential as a function of memory times  $\tau_{\text{o}}$ and $\tau_\varphi$ as well as inertial time  $\tau_{\text{m}}$, from  eqs.(\ref{eq.mfpt_rel_theta},\ref{eq.theta_dw},\ref{eq.w_i_expl},\ref{eq.t_rel}) by using the poles $\omega_k$ of $\tilde{\chi}(\omega)$ in eq.(\ref{eq.chi_t}).}
    \label{fig.mfpt_euler}
\end{figure}

\noindent The relaxation time $\tau_{\text{rel}}$, defined in eq.(13) in the main text, is a polynomial function of the poles $\omega^\mp_k$ of the response function $\tilde{\chi}(\omega)$

\begin{equation}
    \begin{split}
        \tau_{\text{rel}} 
        &=2 \int_{0}^{\infty} (C(t)/C_0)^2 dt\\
        &=\frac{2(k_BT)^2}{C_0^2} \int_{0}^{\infty} \left(  \frac{\sum_k^4 (-1)^k  g^\mp_k }{\underbrace{64  (z_1 - z_2)(z_1 - z_3)(z_2 - z_3) K\aleph^2}_{=\aleph_D} }  e^{i \omega^\mp_k t}    \right)^2 dt\\
        &= \frac{2 i K^2}{ \aleph_D^2} \left[ \frac{(g_1^\mp)^2}{2 \omega_1^\mp } + \frac{(g_2^\mp)^2}{2 \omega_2^\mp } + \frac{(g_3^\mp)^2}{2 \omega_3^\mp } + \frac{(g_4^\mp)^2}{2 \omega_4^\mp } \right] \\
        &+ \frac{4 i K^2}{ \aleph_D^2} \left[ - \frac{g^\mp_1g^\mp_2}{\omega^\mp_1+\omega_2^\mp} + \frac{g^\mp_1g_3^\mp}{\omega_1^\mp+\omega_3^\mp} -  \frac{g^\mp_1g_4^\mp}{\omega_1^\mp+\omega_4^\mp} -  \frac{g^\mp_2g^\mp_3}{\omega_2^\mp+\omega_3^\mp} + \frac{g^\mp_2g_4^\mp}{\omega_2^\mp+\omega_4^\mp} -  \frac{g^\mp_3g_4^\mp}{\omega_3^\mp+\omega_4^\mp} \right],
        \label{eq.t_rel}
    \end{split}
\end{equation}

\noindent where we introduced the shorthand notation 

\begin{equation}
    \aleph_D = 64  (z_1 - z_2)(z_1 - z_3)(z_2 - z_3) K\aleph^2
\end{equation}

\noindent for the denominator. In Figure \ref{fig.mfpt_euler} we plot $\tau_{\text{MFP}} $ using eqs.(\ref{eq.mfpt_rel_theta},\ref{eq.w_i_expl},\ref{eq.t_rel}). Note that $\tau_{\text{rel}}$ in eq.\eqref{eq.t_rel} can be expressed in terms of the roots of the resolvent cubic polynomial $z_k$. Taking into account eq.\eqref{eq.pm_condit}, one can, for example, re-write

\begin{equation}
    \begin{split}
        \omega_1^\mp - \omega_3^\mp &= \left[ \pm (\sqrt{z_1} + \sqrt{z_2} + \sqrt{z_3}) + \frac{i}{2\tau_{\text{o}}} \right] - \left[ \pm (-\sqrt{z_1} + \sqrt{z_2} - \sqrt{z_3}) + \frac{i}{2\tau_{\text{o}}}  \right]\\
        &=\pm 2 (\sqrt{z_1} + \sqrt{z_3})
    \end{split}
\end{equation}

\noindent and

\begin{equation}
    \begin{split}
        \omega_1^\mp + \omega_3^\mp &= \left[\pm(\sqrt{z_1} - \sqrt{z_2} - \sqrt{z_3}) + \frac{i}{2\tau_{\text{o}}} \right] + \left[\pm (-\sqrt{z_1} + \sqrt{z_2} - \sqrt{z_3}] + \frac{i}{2\tau_{\text{o}}}\right]\\
        &= 2 ( \pm \sqrt{z_3} - \frac{i}{2\tau_{\text{o}}})
    \end{split}
\end{equation}

\noindent as well as

\begin{align}
    g^\mp_1 &= \left[\frac{1}{\tau_\varphi^2}+\left(\frac{1}{\tau_{\text{o}}} + i\omega^\mp_1\right)^2\right] (\omega^\mp_2-\omega^\mp_3)(\omega^\mp_2-\omega^\mp_3)(\omega^\mp_2-\omega^\mp_3) \omega^\mp_2\omega^\mp_3\omega^\mp_4\\
    &=\left[\frac{1}{\tau_\varphi^2}+\left(\frac{1}{\tau_{\text{o}}} + i\omega^\mp_1\right)^2\right] 8 (\sqrt{z_1} - \sqrt{z_2} )(\sqrt{z_2} -\sqrt{z_3} )(\sqrt{z_1} -\sqrt{z_3} )\frac{K \aleph^2}{m \omega^\mp_1}.
    \label{eq.g1}
\end{align}

\noindent and
\begin{equation}
    \begin{split}
        g^\mp_1g^\mp_3 &= \left[\frac{1}{\tau_\varphi^2}+\left(\frac{1}{\tau_{\text{o}}} + i\omega^\mp_1\right)^2 \right] \left[\frac{1}{\tau_\varphi^2}+\left(\frac{1}{\tau_{\text{o}}} + i\omega^\mp_3\right)^2\right]      \omega^\mp_1(\omega^\mp_2)^2\omega^\mp_3(\omega^\mp_4)^2\\
        & \times (\omega^\mp_2-\omega^\mp_3)(\omega^\mp_2-\omega^\mp_4)^2(\omega^\mp_3-\omega^\mp_4)(\omega^\mp_1-\omega^\mp_2)(\omega^\mp_1-\omega^\mp_4)\\
         &=f^\mp_1f^\mp_3 \omega^\mp_2\omega^\mp_4 \frac{K}{m}\aleph^2 64 (\sqrt{z_1} -\sqrt{z_2} ) (\sqrt{z_1} -\sqrt{z_3})^2(\sqrt{z_2} -\sqrt{z_3} ) (\sqrt{z_2} +\sqrt{z_3} ) (\sqrt{z_1} +\sqrt{z_2} )\\
        &=f^\mp_1f^\mp_3 \omega^\mp_2\omega^\mp_4 \frac{K}{m}\aleph^2 64 (z_1-z_2)(z_2-z_3)  (\sqrt{z_1} -\sqrt{z_3})^2.
    \end{split}
    \label{eq.g1g3}
\end{equation}

\subsection{Asymptotic discussion of $\lim_{m\rightarrow0} \tau_{\text{rel}}$ }
The relevant coefficients for constructing the poles $z_k = u_k - \frac{d_1}{3 u_k} - \frac{c_2}{3}
$ of $\tilde{\chi}(\omega)$ are

\begin{equation}
\begin{split}
d_1 &= -\frac{1}{m^2} \frac{\left( \frac{\gamma}{ \tau_{\text{o}}} + \frac{\gamma \tau_{\text{o}}}{ \tau_\varphi^{2}} + 2K \right)^2}{192} \\
   &- \frac{1}{m} \left[ \frac{ \frac{\gamma}{ \tau_{\text{o}}} + \frac{\gamma \tau_{\text{o}}}{ \tau_\varphi^{2}} + 2K }{2}\left[\frac{\aleph^2}{24} - \frac{1}{4\tau_{\text{o}}^2} \right] + \frac{K\aleph^2}{4} \right] 
   -\frac{\left(\tau_\varphi^2+\tau_{\text{o}}^2\right)^2}{48 \tau_\varphi^4 \tau_{\text{o}}^4}, \\
d_0 &= \frac{1}{m^3}\frac{\left( \frac{\gamma}{ \tau_{\text{o}}} + \frac{\gamma \tau_{\text{o}}}{ \tau_\varphi^{2}} + 2K \right)^3}{6912}\\
    &+\frac{1}{m^2} \left[ \frac{\frac{\gamma}{ \tau_{\text{o}}} + \frac{\gamma \tau_{\text{o}}}{ \tau_\varphi^{2}} + 2K}{2} \right] \left\{ \left[ \frac{ \frac{\gamma}{ \tau_{\text{o}}} + \frac{\gamma \tau_{\text{o}}}{ \tau_\varphi^{2}} + 2K}{2} \right]\left(\frac{\aleph^2}{288} + \frac{1}{24\tau_{\text{o}}^2} \right) - \frac{K\aleph^2}{24}  \right\}\\
  &+\frac{1}{m} \frac{\left(\tau_\varphi^2+\tau_{\text{o}}^2\right) \left(\gamma \left(-5 \tau_\varphi^4-4 \tau_\varphi^2
   \tau_{\text{o}}^2+\tau_{\text{o}}^4\right)+2 K \tau_\varphi^2 \tau_{\text{o}} \left(\tau_\varphi^2-11
   \tau_{\text{o}}^2\right)\right)}{576 \tau_\varphi^6 \tau_{\text{o}}^5} + \frac{\left(\tau_\varphi^2+\tau_{\text{o}}^2\right)^3}{864
   \tau_\varphi^6 \tau_{\text{o}}^6},
    \end{split}
    \label{eq.d_j_explicit}
\end{equation}

\noindent as well as

\begin{equation}
    \begin{split}
   c_2 &= - \frac{1}{m} \frac{1}{4}\left( \frac{\gamma}{ \tau_{\text{o}}} + \frac{\gamma \tau_{\text{o}}}{ \tau_\varphi^{2}} + 2K \right) + \left( \frac{1}{4\tau_{\text{o}}^{2}} - \frac{1}{2 \tau_\varphi^{2}} \right), 
    \end{split}
\end{equation}

\subsubsection{Constructing roots $z_k$} 

\noindent Using the abbreviations 

\begin{equation}
\begin{aligned}
\lambda_1 &=\frac{\gamma}{ \tau_{\text{o}}} + \frac{\gamma \tau_{\text{o}}}{ \tau_\varphi^{2}} + 2K , & \lambda_2 &= \sqrt{\frac{1}{24} \left(\frac{\gamma}{\tau_\varphi^2\tau_{\text{o}}} + \frac{\gamma}{\tau_{\text{o}}^3} - \frac{2K}{\tau_\varphi^2}\right) }, & \lambda_3 &=\frac{\gamma(\tau_{\text{o}}^2 + \tau_\varphi^2)^2}{\tau_{\text{o}}^3\left[\gamma(\tau_{\text{o}}^2 + \tau_\varphi^2) +2K\tau_{\text{o}} \tau_\varphi^2 \right]^2},
\end{aligned}
\label{eq.abr_lambda}
\end{equation}

\noindent we take the small-mass limit, starting with $u_k$ and $k=1$ of  $\sqrt[3]{+1}  = \epsilon_k=\{1,-\frac{1}{2} \pm i\frac{\sqrt[2]{3}}{2}\}$

\begin{equation}
\begin{split}
    u_1 &= \sqrt[3]{ - \frac{d_0}{2} + \sqrt[2]{ \frac{d_0^2}{4} + \frac{d_1^3}{27}}}\\
    &\approx \sqrt[3]{-1} \left( \frac{\lambda_1}{24m}  - \frac{\lambda_2}{\sqrt{m}} + \frac{\aleph^2}{12} \right).
    \end{split}
\end{equation}

\noindent Since the principal root of $\sqrt[3]{-1}=\{ -\epsilon_3, -1, \epsilon_2 \} $ is $-\epsilon_3$, we construct

\begin{equation}
\begin{split}
\lim_{m\rightarrow 0} u_1 &\approx -\epsilon_3 \left( \frac{\lambda_1}{24 m} - \frac{\lambda_2}{\sqrt{m}} + \frac{\aleph^2}{12} \right) +\mathcal{O}(m^{1/2}) \\ 
\lim_{m\rightarrow 0} u_2 &\approx -1 \left( \frac{\lambda_1}{24 m} - \frac{\lambda_2}{\sqrt{m}}  + \frac{\aleph^2}{12} \right) + \mathcal{O}(m^{1/2}) \\
\lim_{m\rightarrow 0} u_3 &\approx -\epsilon_2 \left( \frac{\lambda_1}{24 m} - \frac{\lambda_2}{\sqrt{m}}  + \frac{\aleph^2}{12} \right) + \mathcal{O}(m^{1/2})
\end{split}    
\label{eq.U_asym}
\end{equation}

\noindent and correspondingly

\begin{equation}
\begin{split}
\frac{d_1}{3u_1} &\approx \frac{\frac{-\lambda_1^2}{192m^2} +\frac{\alpha_1}{m}}{3\left[  (-\epsilon_3) \frac{\lambda_1} {24m}  + \epsilon_3 \frac{\lambda_2}{\sqrt{m}}  \right]}\\
&\approx \epsilon_2 \left( \frac{\lambda_1}{24m} +  \frac{\lambda_2}{\sqrt{m}} +  \frac{\aleph^2}{12} \right) + \mathcal{O}(m^{1/2}).
\end{split}
\label{eq_d1overU_asym}
\end{equation}

\noindent This results in 

\begin{equation}
\begin{split}
\lim_{m\rightarrow 0} \frac{d_1}{3u_1} &\approx  \epsilon_2 \left( \frac{\lambda_1}{24m} + \frac{\lambda_2}{\sqrt{m}} +  \frac{\aleph^2}{12} \right) + \mathcal{O}(m^{1/2}) \\ 
\lim_{m\rightarrow 0} \frac{d_1}{3u_2} &\approx +1 \left( \frac{\lambda_1}{24m} +  \frac{\lambda_2}{\sqrt{m}} + \frac{\aleph^2}{12}  \right) + \mathcal{O}(m^{1/2})\\ 
\lim_{m\rightarrow 0} \frac{d_1}{3u_3} &\approx  \epsilon_3 \left( \frac{\lambda_1}{24m} +  \frac{\lambda_2}{\sqrt{m}}+  \frac{\aleph^2}{12} \right) + \mathcal{O}(m^{1/2})
\end{split}    
\end{equation}

\noindent and for the roots  $z_k = u_k - \frac{d_1}{3 u_k} - \frac{c_2}{3}$ we find 

\begin{equation}
    \begin{split}
      \lim_{m\rightarrow 0}  z_1 &\approx \frac{\lambda_1}{8 m}  - \frac{i\sqrt{3}\lambda_2}{\sqrt{m}}  + \frac{1}{4\tau_\varphi^2} + \mathcal{O}(m^{1/2})\\
      \lim_{m\rightarrow 0}  z_2 &\approx -\frac{1}{4\tau_{\text{o}}^2} + 2\lambda_3 m +  \mathcal{O}(m^{3/2})\\
      \lim_{m\rightarrow 0}  z_3 &\approx \frac{\lambda_1}{8 m}  + \frac{i\sqrt{3\lambda_2}}{\sqrt{m}} + \frac{1}{4\tau_\varphi^2} + \mathcal{O}(m^{1/2})        .
    \end{split}
    \label{eq.z_k_limits}
\end{equation}

\noindent From eqs.\eqref{eq.z_k_limits}, we list further:

\begin{equation}
\begin{aligned}
    z_1-z_2 &\approx \frac{\lambda_1}{8m} + \mathcal{O}(m^{-1/2}) 
    & z_2-z_3 &\approx -\frac{\lambda_1}{8m} + \mathcal{O}(m^{-1/2}) 
    ,\\
    z_1-z_3 &\approx  - 2\frac{i \sqrt{3}\lambda_2}{\sqrt{m}} , & \sqrt{z_1} &\approx \sqrt{\frac{\lambda_1}{8m}} - 4\frac{i \sqrt{3}\lambda_2}{\sqrt{8\lambda_1}} +\mathcal{O}(m^{1/2})\\
    \sqrt{z_2} &\approx \frac{i}{2\tau_{\text{o}}} - 2i\tau_{\text{o}}\lambda_3 m + \mathcal{O}(m^{2}),  & \sqrt{z_3} &\approx \sqrt{\frac{\lambda_1}{8m}} + 4\frac{i \sqrt{3}\lambda_2}{\sqrt{8\lambda_1}}+\mathcal{O}(m^{1/2}),\\
    (\sqrt{z_1} - \sqrt{z_3})^2 &\approx - 24 \frac{\lambda_2^2}{\lambda_1} + \mathcal{O}(m^{1}), & (\sqrt{z_1} + \sqrt{z_3})^2 &\approx \frac{\lambda_1}{2m} + \frac{1}{\tau_\varphi^2} + \frac{\lambda_2^2}{\lambda_1} + \mathcal{O}(m^{1}),\\
    (\sqrt{z_1} - \sqrt{z_2})^2 &\approx  \frac{\lambda_1}{8 m} + \mathcal{O}(m^{-1/2}), & (\sqrt{z_1} + \sqrt{z_2})^2 &\approx \frac{\lambda_1}{8 m} + \mathcal{O}(m^{-1/2}), \\
    (\sqrt{z_2} - \sqrt{z_3})^2 &\approx  \frac{\lambda_1}{8 m} + \mathcal{O}(m^{-1/2}), & (\sqrt{z_2} + \sqrt{z_3})^2 &\approx \frac{\lambda_1}{8 m} + \mathcal{O}(m^{-1/2}).
\end{aligned}
\label{eq.z_approx}
\end{equation}    

\noindent For the expansion of the poles $\omega_k^\mp$, the sign condition in Eq.~\eqref{eq.pm_condit} becomes relevant. As an illustrative example, one finds

\begin{equation}
    \begin{split}
        \omega_1^- &= \sqrt{z_1} + \sqrt{z_2} + \sqrt{z_3} + \frac{i}{2\tau_{\text{o}}} \\
        &\approx \sqrt{\frac{\lambda_1}{8m}}\left( 1 - \frac{4i\sqrt{3m}\lambda_2}{\lambda_1}  \right) + \frac{i}{2\tau_{\text{o}}} + \sqrt{\frac{\lambda_1}{8m}}\left( 1 - \frac{4i\sqrt{3m}\lambda_2}{\lambda_1}  \right)  + \frac{i}{2\tau_{\text{o}}} + \mathcal{O}(\sqrt{m})\\
        &= \frac{i}{\tau_{\text{o}}} + \sqrt{\frac{\lambda_1}{2m}} + \mathcal{O}(\sqrt{m}),
    \end{split}
\end{equation}

\noindent whereas for the opposite sign one obtains

\begin{equation}
    \begin{split}
        \omega_1^+ &=- \left[  \sqrt{z_1} + \sqrt{z_2} + \sqrt{z_3}\right] + \frac{i}{2\tau_{\text{o}}} \\
        &= - \sqrt{\frac{\lambda_1} +{2m}}. +\mathcal{O}(\sqrt{m}).
    \end{split}
\end{equation}

\noindent To state the relevant expansions involving $\omega_k^\mp$ in a compact fashion, we introduce $\alpha^\mp=\{ 0,1 \}$, to write

\begin{equation}
    \begin{aligned}
        \omega_1^\mp &\approx \pm \sqrt{\frac{\lambda_1}{2m}} + \alpha^\mp \frac{i}{\tau_{\text{o}}} + \mathcal{O}(m^{1/2}) , \quad 
        \begin{array}{c}
            \alpha^- = 1, \text{ ( for } \omega^- \text{)}\\
            \alpha^+ = 0, \text{ ( for } \omega^+ \text{)}
        \end{array} \\
        \omega_2^\mp &\approx \mp \frac{\sqrt{24}i\lambda_2}{\sqrt{\lambda_1}} + \alpha^\mp \frac{i}{\tau_{\text{o}}} + \mathcal{O}(m^{1}), \quad 
        \begin{array}{c}
            \alpha^+ = 1 \\
            \alpha^- = 0
        \end{array}\\
        \omega_3^\mp &\approx \mp \sqrt{\frac{\lambda_1}{2m}} + \alpha^\mp \frac{i}{\tau_{\text{o}}} + \mathcal{O}(m^{1/2}), \quad 
        \begin{array}{c}
            \alpha^- = 1 \\
            \alpha^+ = 0
        \end{array} \\
        \omega_4^\mp &\approx \pm \frac{\sqrt{24}i\lambda_2}{\sqrt{\lambda_1}} + \alpha^\mp \frac{i}{\tau_{\text{o}}} + \mathcal{O}(m^{1}), \quad 
        \begin{array}{c}
            \alpha^+ = 1 \\
            \alpha^- = 0
        \end{array}\\
    \end{aligned}
    \label{eq.w_k_lim}
\end{equation}

\noindent and for $f_k^\mp = \frac{1}{\tau_\varphi^2}+\left(\frac{1}{\tau_{\text{o}}} + i\omega^\mp_k\right)^2$
\begin{equation}
    \begin{aligned}
        f_1^\mp &\approx - \frac{\lambda_1}{2m} + \mathcal{O}(m^{0}) \\
        f_3^\mp &\approx - \frac{\lambda_1}{2m} + \mathcal{O}(m^{0}) \\
        f_2^\mp &\approx \frac{1}{\tau_\varphi^2} + \frac{24\lambda_2^2}{\lambda_1} + \alpha^\mp \left( \frac{1}{\tau_{\text{o}}^2}  + \frac{\sqrt{96} \lambda_2}{\sqrt{\lambda_1}\tau_{\text{o}}}\right) + \mathcal{O}(m^{1}), \quad 
        \begin{array}{c}
            \alpha^- = 1 \\
            \alpha^+ = 0
        \end{array}\\
        f_4^\mp &\approx \frac{1}{\tau_\varphi^2} + \frac{24\lambda_2^2}{\lambda_1} + \alpha^\mp \left( \frac{1}{\tau_{\text{o}}^2}  - \frac{\sqrt{96} \lambda_2}{\sqrt{\lambda_1}\tau_{\text{o}}}\right) + \mathcal{O}(m^{1}), \quad 
        \begin{array}{c}
            \alpha^- = 1 \\
            \alpha^+ = 0.
        \end{array}
    \end{aligned}
    \label{eq.f_k_lim}
\end{equation}

\subsubsection{Upper limit for $\lim_{m\rightarrow0} m\cdot\tau_{\text{rel}}$}

\begin{figure}[htbp]
   \centering 
    \includegraphics[height=0.25\textheight]{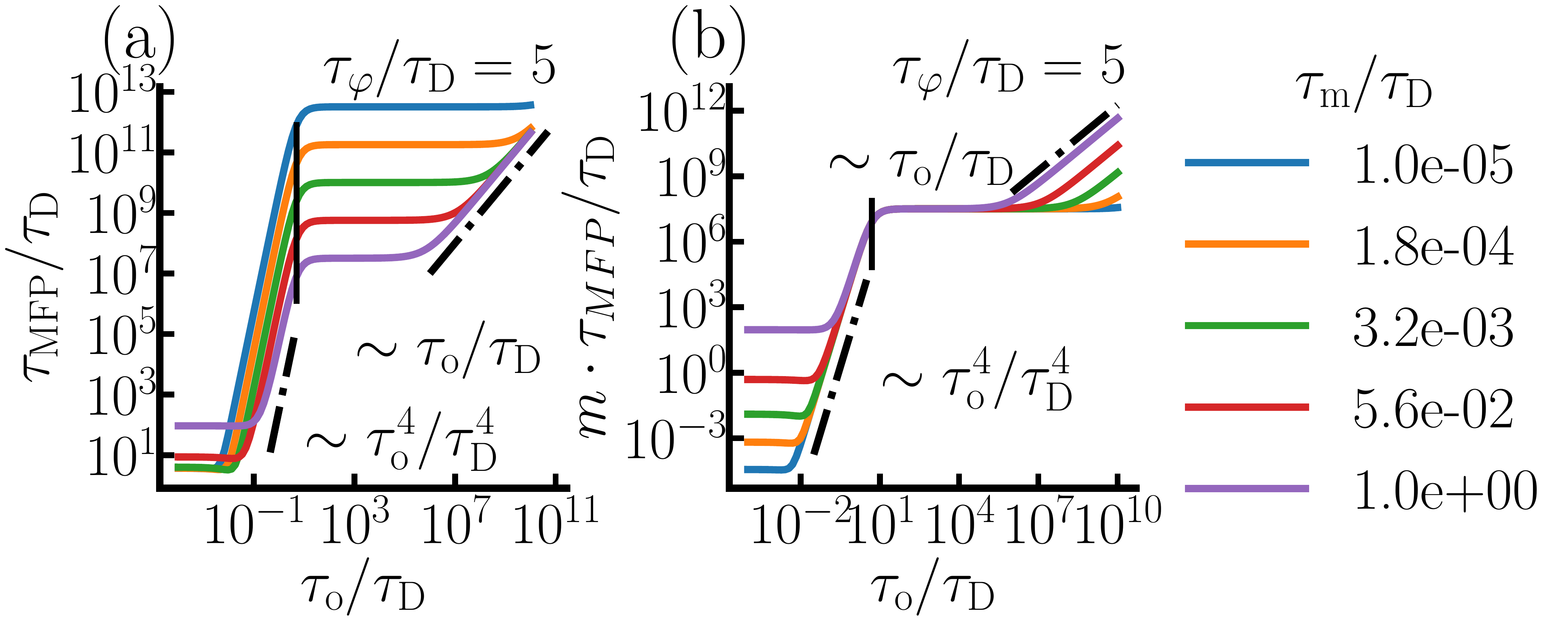}

    \caption{(a) $\tau_{\text{MFP}}$ from eqs.(\ref{eq.mfpt_rel_theta},\ref{eq.theta_dw},\ref{eq.w_i_expl},\ref{eq.t_rel}) as a function of $\tau_{\text{o}}/\tau_{\text{D}}$, with fixed $\tau_\varphi/\tau_{\text{D}}=5$ (vertical bar), (b) $m \cdot \tau_{\text{MFP}}$. Black dash-dotted lines indicate power-law scaling. }
    \label{fig.mfpt_m_sum2}
\end{figure}

From Figure \ref{fig.mfpt_euler} (c), we see that $\tau_{\text{MFP}}$ diverges for small $m$, so we infer that $m \cdot\tau_{\text{MFP}}$ converges for small $m$, see Figure \ref{fig.mfpt_m_sum2}(b).
Next, we derive the leading order behaviour for each term in $\tau_{\text{MFP}}$, eq.\eqref{eq.t_rel}, starting with the quadratic terms

\begin{align}
    2 i K^2  \frac{(g_1^\mp)^2}{2 \omega^\mp_1 \aleph_D^2} &=  2 i K^2 \frac{ (f_1^\mp)^2  (\sqrt{z_1} - \sqrt{z_2} )^2(\sqrt{z_2} -\sqrt{z_3} )^2(\sqrt{z_1} -\sqrt{z_3} )^2}{(\omega_1^\mp)^3 m^2 \left[(z_1-z_2)(z_2-z_3)(z_1-z_3) \right]^2}.
     \label{eq.g1_sq}
\end{align}

\noindent Reading off the $m$-scaling of each factor in eq.\eqref{eq.g1_sq} from eqs.(\ref{eq.z_approx},\ref{eq.w_k_lim},\ref{eq.f_k_lim}), we obtain

\begin{align}
2 i K^2 \frac{(g_1^\mp)^2}{2 \omega^\mp_1 \aleph_D^2} &\sim \frac{m^{-2} m^{-1} m^{-1}m^{0}}{m^{-3/2}m^2 m^{-5}} \sim m^{1/2} \rightarrow 0,
\end{align}

\noindent and for $2 i K^2 \frac{(g_3^\mp)^2}{2 \omega^\mp_3 \aleph_D^2}$ we find the same result. For the next diagonal term, we find 

\begin{equation}
\begin{split}
    2 i K^2 \frac{(g_2^\mp)^2}{2 \omega^\mp_2 \aleph_D^2} &=  2 i K^2 \frac{ (f_2^\mp)^2  (\sqrt{z_1} +\sqrt{z_2} )^2(\sqrt{z_2} -\sqrt{z_3} )^2(\sqrt{z_1} + \sqrt{z_3} )^2}{\omega_2^3 m^2 \left[(z_1-z_2)(z_2-z_3)(z_1-z_3) \right]^2} \\
    &\approx \frac{m^{0} m^{-1} m^{-1}m^{-1}}{m^{0}m^2 m^{-5}} \sim m^{0} ,
     \label{eq.g2_sq}
     \end{split}
\end{equation}

\noindent and  similarly for $2 i K^2 \frac{(g_4^\mp)^2}{2 \omega^\mp_4 \aleph_D^2}$.
Next, we check the off-diagonal terms from  eq.\eqref{eq.t_rel}, for $g^\mp_1g^\mp_3$, rewritten as in eq.\eqref{eq.g1g3}, and after insertions into $ 2 i K^2 \frac{g^\mp_1g^\mp_3}{ (\omega^\mp_1+\omega^\mp_3) \aleph_D^2} $, we find

\begin{equation}
    \begin{split}
         4 i K^2 \frac{g^\mp_1g^\mp_3}{ (\omega^\mp_1+\omega^\mp_3) \aleph_D^2} 
         &=  \frac{i K/m}{32 \aleph^2  } \frac{ f^\mp_1f_3^\mp  \omega^\mp_2\omega^\mp_4(\sqrt{z_1} -\sqrt{z_3} )^2 (\pm \sqrt{z_2}+\frac{i}{2\tau_{\text{o}}})}{ (z_1-z_2)(z_2-z_3)(z_1-z_3)^2  (z_2+\frac{1}{4\tau_{\text{o}}^2}) }\\ 
         &\sim  m^{-1} ,
    \end{split}
    \label{eq.g13_interm}
\end{equation}

\noindent whereas all other off-diagonal terms in eq.\eqref{eq.t_rel} scale with $m^{0}$ or $m^{1/2}$, and therefore do not contribute in the limit $m\rightarrow0$. Therefore, we can state as an upper bound for the relaxation time in the zero mass limit

\begin{equation}
    \lim_{m\rightarrow0} \tau_{\text{rel}} < 4 i K^2  \frac{g^\mp_1g^\mp_3}{ (\omega^\mp_1+\omega^\mp_3) \aleph_D^2} . 
\end{equation}

\noindent With $\lim_{m\rightarrow0}(z_1-z_2)(z_2-z_3)(z_1-z_3)^2 \approx \frac{3}{16}\frac{\lambda_1^2\lambda_2^2}{m^3} + \mathcal{O}(m^{-2})$, and inserting eqs.(\ref{eq.z_approx},\ref{eq.w_k_lim},\ref{eq.f_k_lim}) into eq.\eqref{eq.g13_interm}, we find

\begin{equation} 
    \begin{split}
       \lim_{m\rightarrow0}   4 i K^2 \frac{g^\mp_1g^\mp_3}{ (\omega^\mp_1+\omega^\mp_3) \aleph_D^2} 
         &\approx \frac{i \frac{K}{m} (-24\frac{\lambda_2^2}{\lambda_1})}{32 \aleph^2 \frac{3}{16}\frac{\lambda_1^2\lambda_2^2}{m^3} } \frac{\frac{\lambda_1^2}{4m^2} \omega^\mp_2\omega^\mp_4 (\pm \sqrt{z_2}+\frac{i}{2\tau_{\text{o}}})}{2\lambda_3 m} + \mathcal{O}(m^0)\\
         &=\frac{-i K \omega^\mp_2\omega^\mp_4 (\pm \sqrt{z_2}+\frac{i}{2\tau_{\text{o}}}) }{2 \aleph^2\lambda_1\lambda_3 m} + \mathcal{O}(m^0).
     \end{split}
    \label{eq.g13_interm2}
\end{equation}

\noindent For $\sqrt{z_1z_2z_3}=+\frac{b_1}{8}$, we obtain 

\begin{equation}
    \begin{split}
   \lim_{m\rightarrow0}  4 i K^2 \frac{g^+_1g^+_3}{ (\omega^+_1+\omega^+_3) \aleph_D^2} 
         &\approx \frac{-i K (24\frac{\lambda_2^2}{\lambda_1} - \frac{1}{\tau_{\text{o}}}) (-\frac{i}{\tau_{\text{o}}}) }{2 \aleph^2\lambda_1\lambda_3 m} + \mathcal{O}(m^0)\\
         &=\frac{\frac{K}{\tau_{\text{o}}} (\frac{\lambda_1}{\tau_{\text{o}}^2 - 24\lambda_2^2}) }{2 \aleph^2\lambda_1^2\lambda_3 m}+ \mathcal{O}(m^0)\\
         &= \frac{K^2}{\gamma m } \frac{\tau_{\text{o}}^4 \tau_\varphi^4}{(\tau_{\text{o}}^2 + \tau_\varphi^2)^2} + \mathcal{O}(m^0).
     \end{split}
    \label{eq.g13_fin}
\end{equation}

\noindent The resulting $\frac{\tau_{\text{o}}^4 \tau_\varphi^4}{(\tau_{\text{o}}^2 + \tau_\varphi^2)^2}$-term gives the $\tau_{\text{MFP}} \sim  \tau_j^4$ scaling whenever $\tau_j<\tau_k$ with $\tau_{j,k} \in \{\tau_{\text{o}},\tau_\varphi \}$. This corresponds to the scaling regimes characterized by regimes (III) and (IV) in Figure 4 of the main text. 
Thus, from eqs.(\ref{eq.g13_fin}) we obtain for the relaxation time 

\begin{equation}
   \tau_{\text{rel}} = \frac{1}{m} \left( \frac{K^2}{\gamma } \frac{\tau_{\text{o}}^4 \tau_\varphi^4}{(\tau_{\text{o}}^2 + \tau_\varphi^2)^2}  \right) + \mathcal{O}(m^0). 
   \label{eq.t_rel_asym_fin}
\end{equation}

\noindent In terms of the variables in the Markovian embedding, eq.\eqref{eq.t_rel_asym_fin} scales according to

\begin{equation}
\begin{split}
       \tau_{\text{rel}} &= \frac{1}{m} \left( \frac{K^2}{\gamma } \frac{m_y^2}{k_y^2} \right) + \mathcal{O}(m^0). \\
       &= m  \frac{K^2}{\gamma }  \frac{(m_y/m)^2}{k_y^2}  + \mathcal{O}(m^0).
       \end{split}
   \label{eq.t_rel_asym_fin_inmy}
\end{equation}

\noindent Thus, in the limit \(m_y\to0\), the leading \(\mathcal{O}(1/m)\) contribution vanishes, and with it the quartic scaling with \(\tau_{\mathrm{o}}\) and \(\tau_\varphi\) of the relaxation time.

\section{$\tau_{\text{MFP}}$ for GLE systems with oscillatory-exponential memory kernel in harmonic potential}

In this section, we compare $\tau_{\text{MFP}}$ for an oscillatory-exponential memory kernel in a harmonic well potential, eq.(\ref{eq.hw}), as predicted from theory in section \ref{sec.deriving_tau_rel} with simulation results of the GLE.

\begin{figure*}
\centering
\includegraphics[width=0.99\textwidth]{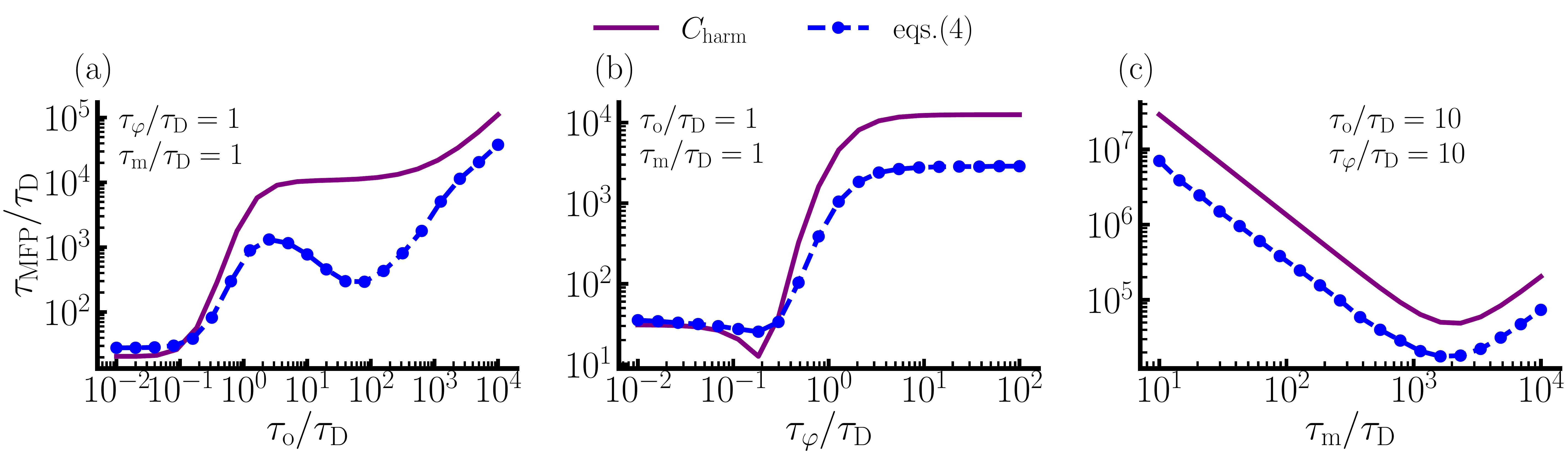} 
\caption{Comparison of barrier-crossing times in a harmonic well potential, eq.(\ref{eq.hw}) obtained from GLE simulations (blue circles with dashed lines), and from analytic theory eq.(\ref{eq.t_rel}), based on $C_{\text{harm}}$ (purple solid curves). (a) $\tau_{\text{MFP}}$ as a function of $\tau_{\text{o}}/\tau_{\text{D}}$ with $\tau_\varphi/\tau_{\text{D}}=1$ and $\tau_{\text{m}}/\tau_{\text{D}}=1$.
(b) $ \tau_{\text{MFP}} $ as a function of $\tau_\varphi/\tau_{\text{D}}$ with $\tau_{\text{o}}/\tau_{\text{D}}=1$ and $\tau_{\text{m}}/\tau_{\text{D}}=1$. (c) $ \tau_{\text{MFP}} $ as a function of $\tau_{\text{m}}/\tau_{\text{D}}$ with $\tau_\varphi/\tau_{\text{D}}=10$ and $\tau_{\text{o}}/\tau_{\text{D}}=10$.}
\label{fig.OSC_MFPT_SIM_ANALYT_HW}
\end{figure*}

 Similar to Figure 5 in the main text, Figure \ref{fig.OSC_MFPT_SIM_ANALYT_HW} shows $\tau_{\text{MFP}}$ predicted from eq.(\ref{eq.t_rel}) using $C_\text{harm}(t)$ given in eq.\eqref{eq.C_t} (purple curves) and $\tau_{\text{MFP}}$ extracted from trajectories $C_\text{harm}(t)$ (blue circles with dashed lines), as functions of the timescale ratios $\tau_{\text{o}}/\tau_\text{D}, \tau_\varphi/\tau_\text{D}, \tau_{\text{m}}/\tau_\text{D} $. The theoretical prediction, eq.(\ref{eq.t_rel}), does not fully agree with the $\tau_{\text{MFP}}$ extracted from trajectories in a harmonic well. These deviations are similar to those observed in a double well potential (Figure 5 in the main text), although the disagreement is smaller in the harmonic well. The persistence of the discrepancy for a harmonic potential shows that it is not a consequence of the harmonic approximation in the theory. Rather, it reflects the limitations of the cumulant expansion of the time-dependent diffusivity leading to eq.(13) of the main text \cite{Netz26}.

\medskip 
\bibliography{refs-si}